\documentclass[english,aps,prr,longbibliography,superscriptaddress,floatfix, notitlepage,reprint]{revtex4-1}

\usepackage[T1]{fontenc}
\usepackage{graphicx}% Include figure files
\usepackage{bbm}% bold math
\usepackage{float} 
\usepackage[usenames,dvipsnames]{xcolor}
\usepackage{amsmath} 
\usepackage[colorlinks=true,citecolor=Cerulean,linkcolor=RubineRed,urlcolor=Cerulean]{hyperref}
\usepackage{stmaryrd}
\usepackage{tabularx}
\usepackage{comment}

\usepackage{mnsymbol}
\usepackage[most]{tcolorbox}
\tcbuselibrary{skins,breakable}

\usepackage{etoolbox}
\usepackage{mathrsfs}

\tcolorboxenvironment{lemma}{
  enhanced,
  colback=blue!5,
  colframe=white,
  boxrule=0pt,
  sharp corners,
  left=6pt,right=6pt,top=6pt,bottom=6pt
}

\tcolorboxenvironment{theorem}{
  enhanced,
  colback=green!5,
  colframe=white,
  boxrule=0pt,
  sharp corners,
  left=6pt,right=6pt,top=6pt,bottom=6pt
}

\begin{document}

\title{Explicit closed-form propagators for time-dependent gain--loss master equations}

\author{L\'eonce Dupays   \href{https://orcid.org/0000-0002-3450-1861}{\includegraphics[scale=0.05]{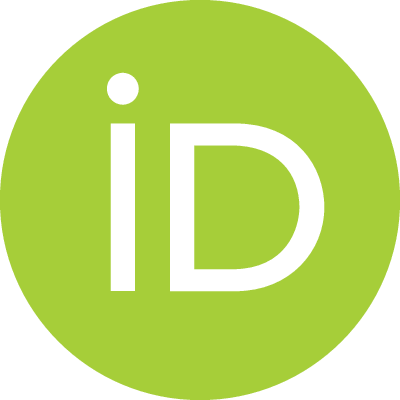}}}
\email{leonce.dupays@kcl.ac.uk}
\affiliation{Department of Mathematics, King’s College London, Strand, London WC2R 2LS, UK}
\begin{abstract}
Time ordering is a central obstacle in driven open-system dynamics governed by master equations with time-dependent rates. We derive explicit closed-form factorizations of the chronological and antichronological propagators for a time-dependent generator of the form $\mathcal{L}(t)=\alpha(t)A+\beta(t)B$, when the commutator $[A,B]$ is a simultaneous eigenoperator of the adjoint actions of $A$ and $B$, namely $[A,[A,B]]=\lambda_{A}[A,B]$ and $[B,[A,B]]=\lambda_{B}[A,B]$. In contrast to the disentangling procedure known as the Wei--Norman decomposition, our construction determines the factorization coefficients directly, without solving a coupled system of nonlinear differential equations. Applying this result to the bosonic gain--loss Lindblad equation, we obtain the exact dynamical map for arbitrary time-dependent gain and loss rates. The gain--loss master equation provides a canonical description of a driven harmonic oscillator coupled to a thermal bath. We benchmark the construction by deriving the time evolution of the Wigner function of an initial Fock state and comparing it with independent phase-space results. We further apply the resulting dynamical map to evaluate the code-space survival probability of a binomial bosonic code under driven gain--loss dynamics. 
\end{abstract}
\maketitle
\section{Introduction}
Driven open quantum systems provide a natural setting for studying nonequilibrium state preparation, transport, thermodynamics, and controlled relaxation \cite{Verstraete2009,Diehl2008,Barreiro2011,Shankar2013, Vacanti2014, Alipour2020,Sieberer2016,Kalthoff2022,Rakovszky2024,Kosloff2013,Benenti2017,Myers2022,Landi2022,Popkov2020,Haddadfarshi2015,Schnell2020,Dupays2020,Dupays2021,Mori2023,Dann2025,Keliri2026}. The reduced dynamics of such systems is generally more involved than in the time-independent case because the drive modifies both the system Hamiltonian and its coupling to the environment. Microscopic master equations have been developed in the adiabatic regime, for periodically driven systems, and beyond the adiabatic limit \cite{Kamleitner2011,Reimer2018,Dann2019,Shavit2019,Wu2022,DiMeglio2024}. Under suitable weak-coupling, short-memory, and secular approximations, the resulting dynamics can be written in time-local Gorini--Kossakowski--Sudarshan--Lindblad (GKSL) form \cite{Gorini1976,Lindblad1976,Stefanini2026},
\begin{align}
\frac{d}{dt}\rho_S(t)=\mathcal{L}(t)\rho_S(t).
\end{align}
Because the generator is explicitly time dependent, $\mathcal{L}(t_1)$ and $\mathcal{L}(t_2)$ need not commute. The corresponding dynamical map therefore takes the time-ordered form
\begin{eqnarray}
V(t,s)\equiv \mathcal{T}_{+}\exp\left[\int_{s}^{t}dt'\mathcal{L}(t')\right],
\end{eqnarray}
where $\mathcal{T}_{+}$ denotes chronological ordering. Obtaining an explicit solution therefore amounts to removing this time ordering.
A standard numerical approach is to approximate the propagator by products over short time intervals, using Trotter--Suzuki formulas and their higher-order generalizations \cite{Trotter1959,Suzuki1976,Suzuki1990,Huyghebaert1990,David2025,Pillay2026}. Such constructions are systematically improvable but remain approximate. Exact analytical factorizations can instead be obtained when the generator possesses a suitable Lie-algebraic structure \cite{Scopa2019}. 
\begin{figure}[!htbp]
    \centering
    \includegraphics[width=\linewidth]{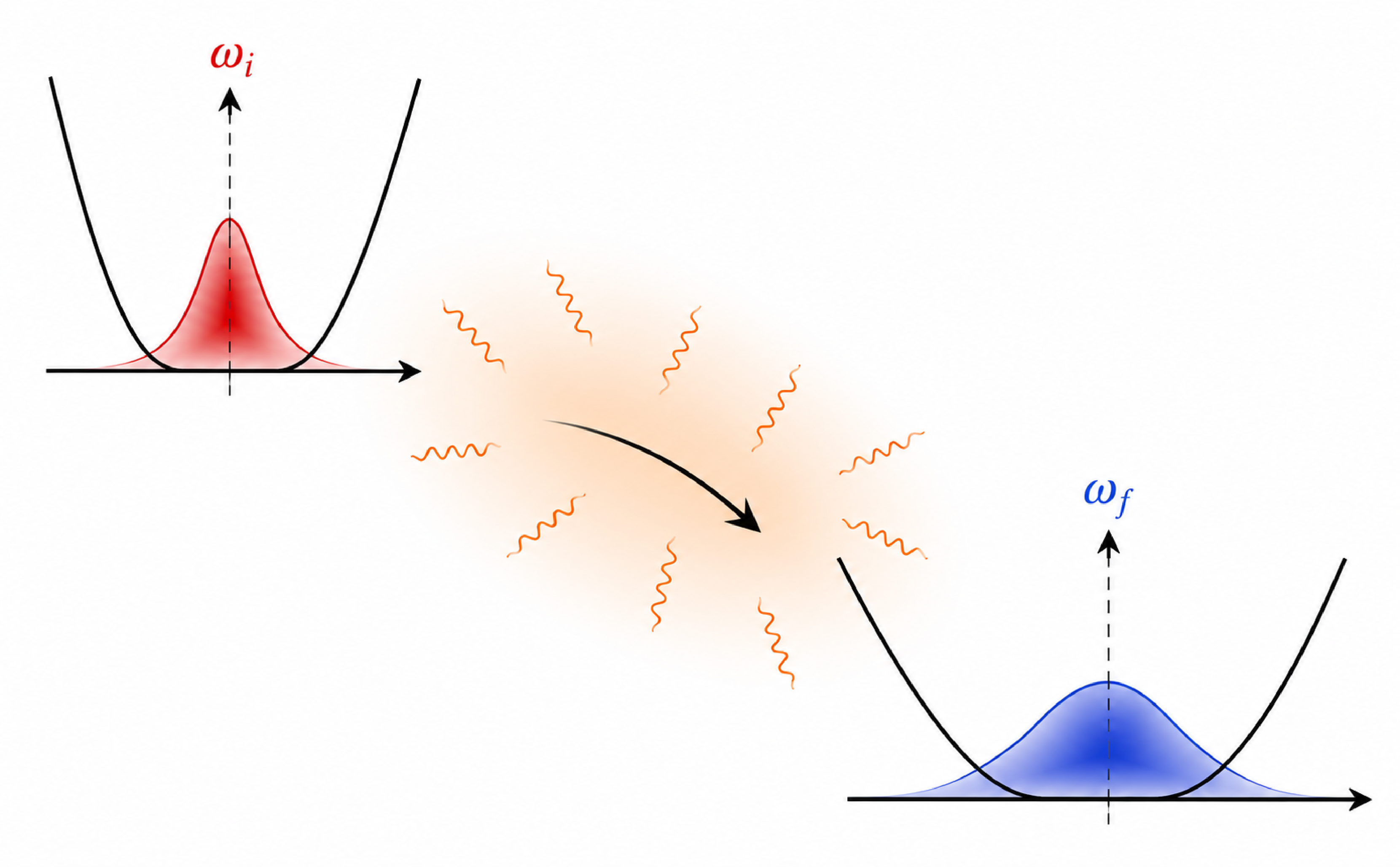}
    \caption{{\bf Possible physical realization of the driven gain--loss dynamics}. A particle is confined in a time-dependent harmonic potential and coupled to a thermal bath. The interplay between driving and dissipation leads to a gain--loss master equation with time-dependent rates \cite{Dann2018}. We provide closed-form expressions for the chronological and antichronological propagators, which can be used to characterize the dynamics of an arbitrary initial state. }
    \label{fig:sketch}
\end{figure}
The Wei--Norman decomposition expresses the associated propagator as an ordered product of exponentials of Lie-algebra generators \cite{Wei1963,Charzynsky2013}, and its scalar coefficients are determined through a coupled system of nonlinear differential equations. Here we focus on an algebraic structure for which the factorization coefficients can be obtained directly. Specifically, we consider
\begin{align}
\mathcal{L}(t)=\alpha(t)A+\beta(t)B, \label{specific_Linblad}
\end{align}
with time-independent generators $A$ and $B$. We focus on pairs of generators whose commutator is a simultaneous eigenoperator of the adjoint actions of the two operators. This property closes all higher nested commutators onto a single operator and allows the time ordering to be removed explicitly. Building on time-dependent operator factorizations \cite{Wilcox1967,Suzuki1985,Suzuki1993}, we derive explicit closed-form decompositions of both the chronological and antichronological propagators for arbitrary time-dependent coefficients $\alpha(t)$ and $\beta(t)$. 

We then apply this result to the bosonic gain--loss Lindblad equation. Its loss and gain dissipators generate a two-dimensional algebra that can be recast in terms of the $su(1,1)$ algebra generators \cite{Fujii2008,Chruscinski2010,Ringel2012,Ringel2013,Scopa2019,Korsch2019,Teuber2020,Lin2025} and satisfy precisely the eigencommutator closure required by our construction. This yields an exact dynamical map for arbitrary time-dependent gain and loss rates, including those arising in a driven harmonic oscillator coupled to a thermal bath. The resulting factorization gives direct access to density-matrix evolution. As applications, we derive the evolution of an initial Fock state and its Wigner function, and benchmark these results against independent calculations based on the corresponding birth--death and phase-space equations. We also derive an analytical expression for the code-space survival probability of a binomial bosonic code under driven gain--loss dynamics.

The paper is organized as follows. In Section~\ref{closed_formula}, we derive the closed-form factorizations of the chronological and antichronological propagators. In Section~\ref{algebra_dissipators}, we apply the construction to the algebra generated by the dissipators of the bosonic gain--loss master equation. In Section~\ref{direct_computation}, we study the evolution of an initial Fock state and its Wigner function, and evaluate the code-space survival probability of a binomial bosonic code. Section~\ref{conclusion} summarizes our results and discusses possible extensions.
\section{Closed-form factorization of time-ordered propagators\label{closed_formula}}
\subsection{Generalized time-ordered factorization}
As a preliminary step, we derive a closed-form decomposition of the propagator generated by the time-dependent operator in Eq.~\eqref{specific_Linblad}. Suzuki derived a general decomposition of the time-ordered exponential generated by the sum of two time-dependent operators \cite{Suzuki1985}. For arbitrary operators $A(s)$ and $B(s)$, the chronological and antichronological propagators can be written as (identity $9$ of Ref.~\cite{Suzuki1985})
\begin{align}
&\mathcal{T}_{+}\exp\left(\int_{t_{0}}^{t}ds\{A(s)+B(s)\}\right)=\mathcal{T}_{+}\exp\left(\int_{t_{0}}^{t}A(s)ds\right)\nonumber\\
&\times\mathcal{T}_{+}\exp\left(\int_{t_{0}}^{t}B(s)ds\right)\mathcal{T}_{+}\exp\left(\int_{t_{0}}^{t}\hat{C}_{+}(s)ds\right),\label{eq:Suzuki_equation_p}\\
&\mathcal{T}_{-}\exp\left(\int_{t_{0}}^{t}ds\{A(s)+B(s)\}\right)=\mathcal{T}_{-}\exp\left(\int_{t_{0}}^{t}\hat{C}_{-}(s)ds\right)\nonumber\\
&\times\mathcal{T}_{-}\exp\left(\int_{t_{0}}^{t}B(s)ds\right)\mathcal{T}_{-}\exp\left(\int_{t_{0}}^{t}A(s)ds\right),\label{eq:Suzuki_equation_m}
\end{align}
Here, $\mathcal{T}_{+}$ and $\mathcal{T}_{-}$ denote chronological and antichronological ordering, respectively, and the corresponding correction operators are given by Eq.~$(3.49)$ in \cite{Suzuki1985}
\begin{align}
\hat{C}_{\pm}(t)&=\mathcal{T}_{\mp}\exp\left(\mp\int_{t_{0}}^{t}ds\ ad_{B(s)}\right)\nonumber\\
&\times \left[\mathcal{T}_{\mp}\exp\left(\mp\int_{t_{0}}^{t}ds \ ad_{A(s)}\right)-1\right]B(t), \label{Suzuki_correction}
\end{align}
where the adjoint action is defined recursively $ad^{0}_{A}=\mathbbm{1}$, $ad^{1}_{A}B=[A,B]$, and $ad^{n+1}_{A}B=[A,ad^{n}_{A}B]$, with the commutator convention $[A,B]=AB-BA$. Although Eqs.~\eqref{eq:Suzuki_equation_p}-\eqref{eq:Suzuki_equation_m} are exact, the correction term in Eq.~\eqref{Suzuki_correction} generally contains further time-ordered exponentials of adjoint actions. The decomposition therefore does not, by itself, provide an explicit finite factorization. 

We now specialize Suzuki's identity to  $A(s)=\alpha(s)A$ and $B(s)=\beta(s)B$, with $A$ and $B$ time-independent. Under the algebraic conditions 
\begin{align}
[A,[A,B]]&=\lambda_A[A,B],&
[B,[A,B]]&=\lambda_B[A,B],\label{closed_algebra_rules}
\end{align}
the correction operator $\hat{C}_{\pm}(t)$ becomes proportional to the commutator $[A,B]$ at all times, as shown in Appendix~\ref{time_ordering}. The resulting chronological and antichronological propagators are given by
\begin{align}
&\mathcal{T}_{+}\exp\left\{\int_{t_{0}}^{t}ds [\alpha(s)A+\beta(s)B]\right\}\nonumber\\
&=e^{\Gamma_{\alpha}A}e^{\Gamma_{\beta}B}e^{\Delta_{+}[A,B]}=e^{\Gamma_{\alpha}A}e^{\Delta_{+}e^{\Gamma_{\beta} \lambda_{B}}[A,B]}e^{\Gamma_{\beta}B}.\label{eq:main_result}\\
&\mathcal{T}_{-}\exp\left\{\int_{t_{0}}^{t}ds [\alpha(s)A+\beta(s)B]\right\}\nonumber\\
&=e^{\Delta_{-}[A,B]}e^{\Gamma_{\beta}B}e^{\Gamma_{\alpha}A}=e^{\Gamma_{\beta}B}e^{\Delta_{-}e^{-\Gamma_{\beta}\lambda_{B}}[A,B]}e^{\Gamma_{\alpha}A},\label{eq_anti_chrono}
\end{align}
with $\Gamma_{\alpha}(t,t_{0})=\int_{t_{0}}^{t}ds\ \alpha(s)$ and $\Gamma_{\beta}(t,t_{0})=\int_{t_{0}}^{t}ds\ \beta(s)$ denoted for brevity by $\Gamma_{\alpha}$ and $\Gamma_{\beta}$. If $\lambda_{A} \neq 0$
{\small
\begin{align}
\Delta_{\pm}(t,t_{0};\lambda_{A},\lambda_{B})&=\int_{t_{0}}^{t}ds\;\frac{e^{\mp \lambda_{A}\Gamma_{\alpha}(s,t_{0})}-1}{\lambda_{A}}e^{\mp \lambda_{B}\Gamma_{\beta}(s,t_{0})}\beta(s),
\end{align}
}
For $\lambda_{A}=0$ and $\lambda_{B}\neq 0$
\begin{align}
\Delta_{\pm}(t,t_{0};0,\lambda_{B})&=\mp \int_{t_{0}}^{t}ds\;\beta(s)\Gamma_{\alpha}(s,t_{0})e^{\mp \lambda_{B}\Gamma_{\beta}(s,t_{0})}. \label{eq:delta_zero}
\end{align}
Finally, if $\lambda_{A}=\lambda_{B}=0$
\begin{align}
\Delta_{\pm}(t,t_{0};0,0)=\mp \int_{t_{0}}^{t}ds\;\beta(s)\Gamma_{\alpha}(s,t_{0}).
\end{align}
The key advantage of Eqs.~\eqref{eq:main_result} and \eqref{eq_anti_chrono} is that they solve the time-ordering problem explicitly for arbitrary coefficients $\alpha(t)$ and $\beta(t)$, without requiring the solution of a nonlinear system of differential equations \cite{Scopa2018,Scopa2019}. To our knowledge, this closed form for propagators under the commutation relations of Eq.~\eqref{closed_algebra_rules} has not been given previously. Hence, the novelty relies on the closed integral form of the solution. 
\subsection{Time-independent limit}
An important special case is the time-independent limit. For constant coefficients $\alpha(s)=\alpha$ and $\beta(s)=\beta$, time ordering becomes unnecessary and the propagator reduces to (integrated from $0$ to $t$)
\begin{align}
\exp\left[(\alpha A+\beta B)t\right]&=e^{\alpha t A}e^{\beta t B}e^{\bar{\Delta}_{+}[A,B]}\nonumber\\
&=e^{\alpha t A}e^{\bar{\Delta}_{+}e^{\beta t\lambda_{B}}[A,B]}e^{\beta t B}\\
&=e^{\bar{\Delta}_{-}[A,B]}e^{\beta tB}e^{\alpha tA}\nonumber\\
&=e^{\beta tB}e^{\bar{\Delta}_{-}e^{-\beta t\lambda_{B}}[A,B]}e^{\alpha tA},
\end{align}
with (for $\lambda_{A}\neq0,\lambda_{B}\neq0)$
\begin{align}
\bar{\Delta}_{\pm}(t,0;\lambda_{A},\lambda_{B})&=\mp \frac{1-e^{\mp t\beta \lambda_{B}}}{\lambda_{A}\lambda_{B}}\mp\beta\frac{e^{\mp t(\alpha\lambda_{A}+\beta\lambda_{B})}-1}{(\alpha\lambda_{A}+\beta\lambda_{B})\lambda_{A}}.
\end{align}
The case $\alpha=\beta=t=1$, with $\lambda_{A}=u,\lambda_{B}=-v$, recovers the expressions of \cite{Dupays2023}, provided that $\lambda_{A}\lambda_{B}\neq 0$ and $\lambda_{A}\neq-\lambda_{B}$. Special limiting cases are treated in \cite{Dupays2023}. 
\subsection{Reduction to a vanishing eigenvalue \label{reduction_vanishing_eigenvalue}}
When $\lambda_{A}\lambda_{B}\neq 0$, it is always possible to recast the dynamical generator so that it takes the form of the expression in Eq.~\eqref{eq:delta_zero}. Indeed, consider the change of variables
\begin{align}
\alpha(s)A+\beta(s)B=\tilde{\alpha}(s)\tilde{A}+\tilde{\beta}(s)\tilde{B},
\end{align}
with $\tilde{A}=\lambda_{B}A-\lambda_{A}B$, $\tilde{B}=\lambda_{B}A+\lambda_{A}B$ and
\begin{align}
\tilde{\alpha}(s)=\frac{1}{2}\left(\frac{\alpha(s)}{\lambda_{B}}-\frac{\beta(s)}{\lambda_{A}}\right), &\ \tilde{\beta}(s)=\frac{1}{2}\left(\frac{\alpha(s)}{\lambda_{B}}+\frac{\beta(s)}{\lambda_{A}}\right).
\end{align}
By construction, the transformed operators satisfy 
\begin{align}
[\tilde{A},[\tilde{A},\tilde{B}]]&=\tilde{\lambda}_{A}[\tilde{A},\tilde{B}]=0,\\
[\tilde{B},[\tilde{A},\tilde{B}]]&=\tilde{\lambda}_{B}[\tilde{A},\tilde{B}],
\end{align}
with $\tilde{\lambda}_{A}=0$ and $\tilde{\lambda}_{B}=2\lambda_{A}\lambda_{B}$. However, this reformulation does not obviously simplify the determination of the dynamics, since one still needs to apply the operators $\tilde{A}$ and $\tilde{B}$ to the initial state, which need not be an eigenoperator of either $\tilde{A}$ or $\tilde{B}$.

\section{ Exact propagator for the bosonic gain--loss dynamics \label{algebra_dissipators}}
Beyond their intrinsic algebraic interest, Eqs.~\eqref{eq:main_result} and \eqref{eq_anti_chrono} apply directly to the Lie algebra generated by the dissipators of the gain--loss Lindblad master equation. In Sec.~\ref{algebra_closure}, we show that the dissipators of the gain--loss master equation satisfy the required algebraic closure. In Sec.~\ref{generator_dynamics}, we derive the dynamical propagator. 

\subsection{Algebraic closure of the gain--loss dissipators \label{algebra_closure} }
Consider an open system described by the gain--loss master equation
\begin{align}
\frac{d}{dt}\rho_{t}&=\gamma_{-}(t)\mathcal{D}_{a}(\rho_{t})+\gamma_{+}(t)\mathcal{D}_{a^{\dagger}}(\rho_{t}),\label{Lindblad_master_equation}
\end{align}
with the annihilation and creation dissipators \cite{breuer2002}
\begin{align}
\mathcal{D}_{a}(\bullet)&=a\bullet a^{\dagger}-\frac{1}{2}\{a^{\dagger}a,\bullet\},\\
\mathcal{D}_{a^{\dagger}}(\bullet)&=a^{\dagger}\bullet a-\frac{1}{2}\{a a^{\dagger},\bullet\},
\end{align}
and the anticommutator $\{A,B\}=AB+BA$. We do not consider a Hamiltonian term because we work in the interaction picture, similarly to the master equation construction of Ref.~\cite{Dann2018}. However, adding a harmonic oscillator Hamiltonian would not affect the dynamics because it commutes with the dissipators. To expose the relevant Lie algebra, we first express the master equation in a vectorized form and then identify the $su(1,1)$ generators underlying its dissipators. 

Consider the vectorization procedure with the linear map
\begin{eqnarray}
{\rm vec}\left[|\Psi\rangle \langle \phi |\right] \rightarrow  |\Psi \rangle \otimes |\phi \rangle^{*} \equiv |\Psi,\phi \rrangle , \quad \forall~ \Psi,\phi \in \mathcal{H}_{s}.
\end{eqnarray}
where $^{*}$ denotes the complex conjugation, $\otimes$ the tensor product and $\mathcal{H}_{s}$ the system Hilbert space. This corresponds to stacking the rows of the density matrix into a column vector, which is the vectorization convention introduced in \cite{Gyamfi_2020}. With $|\rho_{t}\rrangle$ the vectorized density matrix and $\mathbbm{L}(t)$ the vectorized Lindbladian, Eq.~\eqref{Lindblad_master_equation} can be written as
\begin{align}
\frac{d}{dt}|\rho_{t}\rrangle&=\mathbbm{L}(t)|\rho_{t}\rrangle,\\
\mathbbm{L}(t)&=\gamma_{-}(t)\mathbbm{D}_{a}+\gamma_{+}(t)\mathbbm{D}_{a^{\dagger}},\label{quantum_optical_lindbladian}
\end{align}
with the vectorized dissipator $\mathbbm{D}_{a}=|\mathcal{D}_{a}(\bullet)\rrangle$ and $\mathbbm{D}_{a^{\dagger}}=|\mathcal{D}_{a^{\dagger}}(\bullet)\rrangle$. For a bosonic mode with ladder operators satisfying $[a,a^{\dagger}]=\mathbbm{1}$, the vectorized loss and gain dissipators are
\begin{eqnarray}
\mathbbm{D}_{a}&=&a\otimes a -\frac{1}{2}a^{\dagger}a \otimes \mathbbm{1}-\frac{1}{2}\mathbbm{1}\otimes a^{\dagger}a \label{dissipator_annihilation},\\
\mathbbm{D}_{a^{\dagger}}&=&a^{\dagger}\otimes a^{\dagger} -\frac{1}{2} aa^{\dagger}\otimes \mathbbm{1}-\frac{1}{2}\mathbbm{1}\otimes aa^{\dagger}.\label{dissipator_creation}
\end{eqnarray}
Here, we have used the Fock basis identities $(a^{*})=a$ and $(a^{\dagger})^{*}=a^{\dagger}$. These dissipators are generated by the $su(1,1)$ algebra operators 
\begin{align}
K_{0}&=\frac{1}{2}(a^{\dagger}a\otimes \mathbbm{1}+\mathbbm{1}\otimes a^{\dagger}a+\mathbbm{1}\otimes \mathbbm{1})\label{eq:K0}\\
&=\frac{1}{2}(a^{\dagger}a\otimes \mathbbm{1}+\mathbbm{1}\otimes a a^{\dagger}),\nonumber\\
K_{+}&=a^{\dagger}\otimes a^{\dagger},\label{eq:K+}\\
K_{-}&=a\otimes a,\label{eq:K-}
\end{align}
that satisfy the commutation relations $[K_{0},K_{\pm}]=\pm K_{\pm}$ and $[K_{-},K_{+}]=2K_{0}$. As a consequence, the dissipators can be expressed as
\begin{align}
\mathbbm{D}_{a}&=K_{-}-K_{0}+\frac{1}{2}\mathbbm{1}\otimes \mathbbm{1},\\
\mathbbm{D}_{a^{\dagger}}&=K_{+}-K_{0}-\frac{1}{2}\mathbbm{1}\otimes \mathbbm{1}.
\end{align}
It follows directly that the dissipators of annihilation and creation satisfy the commutation relation
\begin{eqnarray}
[\mathbbm{D}_{a},\mathbbm{D}_{a^{\dagger}}]=-(\mathbbm{D}_{a}+\mathbbm{D}_{a^{\dagger}}),\label{closure_relation_algebra} 
\end{eqnarray}
such that 
\begin{align}
[\mathbbm{D}_{a},[\mathbbm{D}_{a},\mathbbm{D}_{a^{\dagger}}]]&=-[\mathbbm{D}_{a},\mathbbm{D}_{a^{\dagger}}],\\
[\mathbbm{D}_{a^{\dagger}},[\mathbbm{D}_{a},\mathbbm{D}_{a^{\dagger}}]]&=[\mathbbm{D}_{a},\mathbbm{D}_{a^{\dagger}}]. 
\end{align}
The closure relation Eq.~\eqref{closure_relation_algebra} was also noticed in \cite{Scopa2018}. We have thus shown that the Schrödinger-picture dissipators generate an eigencommutator Lie algebra of the form considered in Sec.~\ref{closed_formula}. This is also true for the adjoint dissipators of the gain--loss master equation. However, we do not treat this case here as the Hermitian conjugate of the chronological disentangled propagator directly gives the adjoint propagator. In the next section, we apply this result to obtain an explicit closed-form expression for the dynamical map of the time-dependent gain--loss Lindblad equation.
\subsection{Exact solution for the dynamical propagator\label{generator_dynamics}}
The disentangling formula can be used to express the dynamics generated by the driven gain-loss master equation. We first decompose the dynamical map in terms of the dissipators associated with the annihilation and creation operators. We then show that the resulting expression can be recast in terms of the generators of the $su(1,1)$ algebra, as discussed further in Appendix~\ref{closed_form_dynamical_map}. The propagator reads
\begin{align}
V(t,t_{0})&=\mathcal{T}_{+}\exp\left\{\int_{t_{0}}^{t}ds [\gamma_{+}(s)\mathbbm{D}_{a^{\dagger}}+\gamma_{-}(s)\mathbbm{D}_{a}]\right\}\nonumber\\
&=e^{\Gamma_{+}(t,t_{0})\mathbbm{D}_{a^{\dagger}}}e^{\Gamma_{-}(t,t_{0})\mathbbm{D}_{a}}e^{\Delta_{+}(t,t_{0})[\mathbbm{D}_{a^{\dagger}},\mathbbm{D}_{a}]}\\
&=e^{\Gamma_{+}(t,t_{0})\mathbbm{D}_{a^{\dagger}}}e^{\Delta_{+}(t,t_{0})e^{-\Gamma_{-}(t,t_{0})}[\mathbbm{D}_{a^{\dagger}},\mathbbm{D}_{a}]}e^{\Gamma_{-}(t,t_{0})\mathbbm{D}_{a}},\label{intermed_equation}
\end{align}
with the dynamical coefficients
\begin{align}
\Gamma_{\pm}(t,t_{0})&=\int_{t_{0}}^{t}ds\;\gamma_{\pm}(s),\\
\Delta_{+}(t,t_{0};1,-1)&=\int_{t_{0}}^{t}ds\;(e^{-\Gamma_{+}(s,t_{0})}-1)e^{\Gamma_{-}(s,t_{0})}\gamma_{-}(s). \label{delta_plus}
\end{align}
Interestingly, this decomposition is valid for arbitrary real time-dependent gain and loss rates $\gamma_{\pm}(t)$, including negative rates. For practical calculations, it is convenient to recast the propagator in another operator basis for evolving initial states. The dynamical map can then be factorized as
\begin{align}
V(t,t_{0})&=e^{\Phi_{+}\mathbbm{D}_{a^{\dagger}}}e^{\Phi_{-}\mathbbm{D}_{a}}\label{dissipators_decomposition}\\
&=\mathcal{N}e^{g_{+}K_{+}}e^{g_{0}K_{0}}e^{g_{-}K_{-}},\label{SU11_decomposition}
\end{align}
with the norm $\mathcal{N}=e^{\frac{\Gamma_{-}-\Gamma_{+}}{2}}$ and coefficients, for $Q\neq 0$
\begin{alignat}{2}
\Phi_{+}&=\Gamma_{+}+\ln[Q],&\quad\Phi_{-}&=\Gamma_{-}+\ln[Q],\label{Phi_-}\\
Q&=\Delta_{+}(t,t_{0})e^{-\Gamma_{-}}+1,&\quad e^{g_{0}}&=\frac{e^{-(\Gamma_{+}+\Gamma_{-})}}{Q^{2}},\label{g0}\\
g_{+}&=1-\frac{e^{-\Gamma_{+}}}{Q},&\quad g_{-}&=1-\frac{e^{-\Gamma_{-}}}{Q},\label{g_-}
\end{alignat}
where, for brevity, we write $\Gamma_{\pm}\equiv \Gamma_{\pm}(t,t_{0})$ and $Q\equiv Q(t,t_{0})$. The further factorizations in Eq.~\eqref{dissipators_decomposition}-\eqref{SU11_decomposition} require additional care because of the denominator $Q$. For Markovian rates \cite{breuer2002}($\gamma_{\pm}(t)\geq 0$ for all $t\geq0$), the instantaneous generator $\mathbbm{L}(t)$ is of GKSL form at every time. The corresponding evolution is therefore completely positive-divisible and the dynamical map generated from the initial time to any later time is completely positive and trace-preserving (CPTP) \cite{Rivas_2012,Chruscinski2012}. Hence the quantity $Q>0$ is strictly positive and the logarithm remains real at finite time. In the presence of negative rates, the decomposition remains valid, but the coefficients $\Phi_{\pm}$ may acquire an imaginary branch or be singular at $\Delta_{+}(t,t_{0};1,-1)=-e^{\Gamma_{-}}$. The singularity can appear at finite time, so that the choice of negative rates requires a careful study of $Q$ that is protocol dependent. Such a singularity concerns the chosen factorization coordinates and does not necessarily imply a singularity of the dynamical propagator itself. In particular, there is no logarithm involved in the original decomposition basis Eq.~\eqref{intermed_equation}. Note that the $su(1,1)$ decomposition has real coefficients even if $Q$ takes negative values. The coefficients $g_{+}(t),g_{0}(t),g_{-}(t)$ of the $su(1,1)$ decomposition in Eq.~\eqref{SU11_decomposition} can also be obtained through the Wei--Norman decomposition. It consists of equating the coefficients in front of the generators $K_{0},K_{+},K_{-}$ of the logarithmic derivative of the propagator $\dot{V}V^{-1}$ with the coefficients of the Lindbladian expressed in the $su(1,1)$ basis. One obtains a set of differential equations called the Wei--Norman system
\begin{align}
\dot{g}_{+}&=\gamma_{+}-(\gamma_{+}+\gamma_{-})g_{+}+\gamma_{-}g^{2}_{+},\label{WeiNorman_g+}\\
\dot{g}_{0}&=-(\gamma_{+}+\gamma_{-})+2\gamma_{-}g_{+},\\
\dot{g}_{-}&=e^{g_{0}}\gamma_{-}.\label{WeiNorman_g-}
\end{align}
Our solution is an exact solution to the Wei--Norman system, as detailed in Appendix~\ref{Wei_Norman_SU11}. It is worth noting that the decomposition of the dynamics is not unique. It can be expressed in terms of Lindblad dissipators or in terms of the $su(1,1)$ generators as shown in Eqs.~\eqref{dissipators_decomposition}--\eqref{SU11_decomposition}. The most convenient decomposition to use depends on the initial state. The decomposition in terms of Lindblad dissipators is particularly convenient for describing the evolution of operators that are simultaneous eigenoperators of the dissipators, including the annihilation and creation operators and their powers. By contrast, the $su(1,1)$ decomposition is especially useful for coherent states, generalized coherent states, and Fock states. 
\subsection{Closed-form solution and link to the time-dependent harmonic oscillator}
The Wei--Norman system in Eq.~\eqref{WeiNorman_g+}-\eqref{WeiNorman_g-} admits a closed-form solution in the sense that $g_{+}(t),g_{-}(t),g_{0}(t)$ can be determined at any time from the functional form of $\gamma_{\pm}(t)$. This is generally not possible for an arbitrary time-dependent $SU(1,1)$ problem. For instance, if one considers the time-dependent harmonic oscillator, closed-form expressions for the coefficients $g_{+}(t),g_{-}(t),g_{0}(t)$ after a Wei--Norman decomposition of the dynamics cannot be found for an arbitrary driving frequency. Although the Wei--Norman procedure applies to the time-dependent harmonic oscillator, the solution with arbitrary frequency $\omega(t)$ is obtained by finding two linearly independent solutions of the differential equation $\ddot{u}+\omega^{2}(t)u=0$, which can be explicitly determined only for certain classes of driving protocols \cite{Pinney1950}. This difference can be understood from the underlying Lie-algebraic structure. The gain--loss dissipators span a two-dimensional solvable Lie algebra ($\mathbbm{D}_{a},\mathbbm{D}_{a^{\dagger}}$). This is further illustrated by Eq.~\eqref{dissipators_decomposition}, which factorizes the dynamics in terms of only two generators. By contrast, for the time-dependent harmonic oscillator, an arbitrary driving explores the full three-dimensional non-solvable algebra. 

\section{Applications \label{direct_computation}}
We now illustrate the usefulness of the exact propagator through several applications. In Sec.~\ref{exp_setup}, we introduce an example of a physical setting for gain--loss dynamics with time-dependent rates: a driven harmonic oscillator coupled to a thermal bath. In Sec.~\ref{evolution_Fock_state}, we determine the evolution of the Wigner function of an initial Fock state and compare it with independent derivations, providing a benchmark of the analytical construction. Finally, in Sec.~\ref{bosonic_code}, we use the propagator to derive the code-space survival probability of a binomial bosonic code.

\subsection{Physical realization \label{exp_setup}}
A natural physical setting for the gain--loss master equation is a trapped particle coupled to a thermal bath. Consider a particle immersed in a thermal bath while confined by a time-dependent harmonic potential. The system Hamiltonian reads
\begin{align}
H_{S}(t)
=
\frac{\hat{p}^{2}}{2m}
+
\frac{1}{2}m\omega^{2}(t)\hat{x}^{2},
\end{align}
where $\hat{x}$ and $\hat{p}$ are the position and momentum operators, respectively, $m$ is the mass of the particle and $\omega(t)$ is the time-dependent frequency. The oscillator is weakly coupled to a bosonic thermal reservoir with an Ohmic spectral density through an interaction that is linear in the oscillator quadratures. Following the construction of Refs.~\cite{Dann2018,Dann2019}, the reduced master equation is obtained by first solving the unitary
dynamics generated by $H_S(t)$ and subsequently transforming the
system--bath coupling operator in the interaction
picture. In this interaction picture, the effect of the driving is encoded in a dressed transition frequency and in a time-independent annihilation operator, while the associated gain and loss rates remain explicitly time dependent. The construction of Refs.~\cite{Dann2018,Dann2019} assumes driving through a time-dependent oscillator frequency governed by a constant adiabatic-parameter protocol
\begin{align}
\omega(t)=\frac{\omega_{0}}{1-\xi \omega_{0} t}, \qquad \xi=\frac{\dot{\omega}(t)}{\omega^{2}(t)}=\mathrm{const}.\label{eq_frequency}
\end{align}
The dressed transition frequency is 
\begin{align}
\Omega(t)=\frac{\sqrt{4-\xi^{2}}}{2}\,\omega(t).
\end{align}
The resulting master equation in the interaction picture is of the form
\begin{align}
\dot{\rho}_{t}
=
\gamma_{-}(t)\,\mathcal{D}_{a}(\rho_{t})
+
\gamma_{+}(t)\,\mathcal{D}_{a^\dagger}(\rho_{t}),
\end{align}
where the rates satisfy the thermal detailed-balance condition
\begin{align}
\gamma_{-}(t)=\gamma_{+}(t) e^{\hbar\Omega(t)/(k_{B}T)}=\Gamma_{0}\,\Omega(t)\,[\bar n_{\rm th}(t)+1],\label{eq_gamma}
\end{align}
with $\Gamma_{0}$ the overall dissipation strength and the Bose--Einstein occupation number evaluated at the dressed transition
frequency
\begin{align}
\overline{n}_{\mathrm{th}}(t)
=
\frac{1}{e^{\hbar\Omega(t)/(k_{B}T)}-1}.\label{occupation_number}
\end{align}
\begin{figure*}[t]
    \centering
    \includegraphics[width=0.8\linewidth]{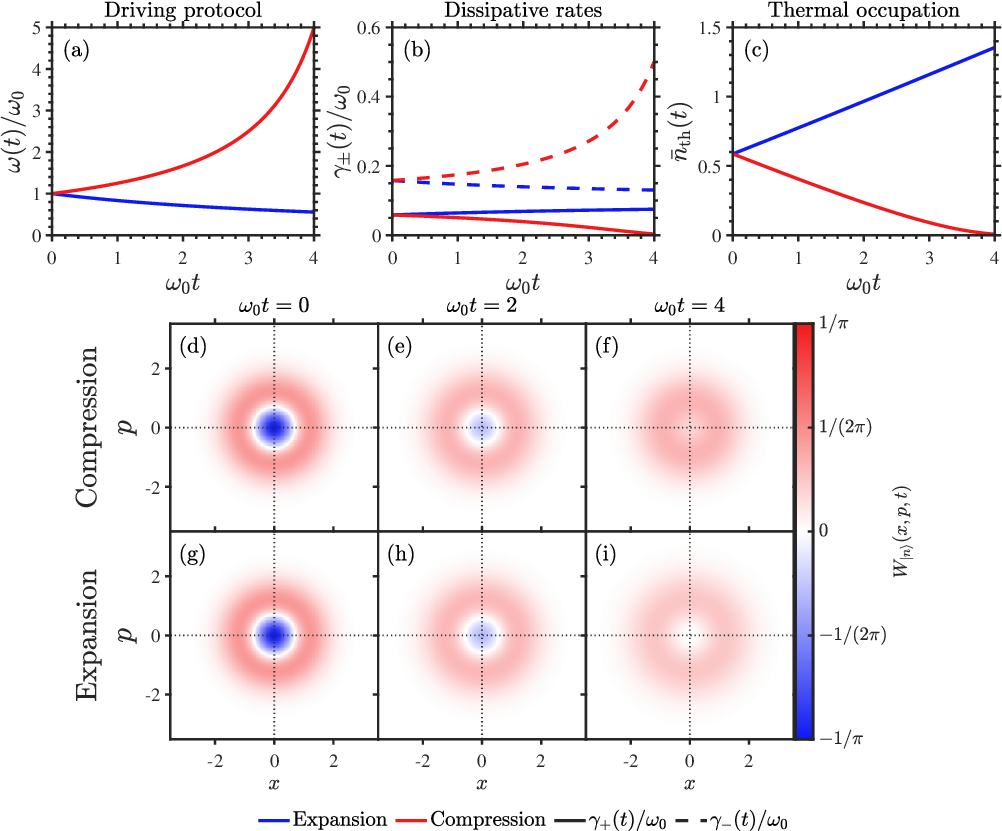}
    \caption{{\bf Driven dissipative dynamics for expansion ($\xi=-0.20$, blue) and compression ($\xi=+0.20$, red) protocols and associated Wigner functions.} The first row shows (a) the oscillator frequency $\omega(t)/\omega_{0}$ given by Eq.~\eqref{eq_frequency}, (b) the excitation and relaxation rates $\gamma_{+}(t)/\omega_{0}$ and $\gamma_{-}(t)/\omega_{0}$ given by Eq.~\eqref{eq_gamma}, and (c) the instantaneous thermal occupation $\bar n_{\rm th}(t)$ given by Eq.~\eqref{occupation_number}. In panel (b), solid and dashed lines denote $\gamma_{+}(t)$ and $\gamma_{-}(t)$, respectively. The second and third rows show the Wigner function of the initial Fock state $|n=1\rangle$ at $\omega_{0}t=0$, $2$, and $4$ for the compression and expansion protocols, respectively. Red and blue regions indicate positive and negative Wigner values. We use the dimensionless parameters $\omega_{0}=1$, $|\xi|=0.20$, $T=1 $, $\Gamma_{0}=0.1$, and $\hbar=k_{B}=1$.}
\label{fig:Wigner_function_evolution}
\end{figure*}
\subsection{Wigner-function evolution of an initial Fock state\label{evolution_Fock_state}}
As an illustration, we consider the evolution of the Wigner function associated with an initial Fock state, $\rho(0)=|n\rangle\langle n|$. For $n>0$, Fock states are non-Gaussian and exhibit genuinely nonclassical features, most notably the negativity of their Wigner function. Their evolution in a thermal environment therefore provides a direct way to visualize the gradual loss of nonclassicality as dissipation suppresses these negative regions. A second motivation for considering Fock states is that their Wigner-function dynamics can be derived independently from the corresponding phase-space evolution equation. This provides an independent benchmark for the exact propagator obtained above. As derived in Appendix~\ref{evolution_Fock_state}, the dynamics modifies both the radial dependence and the amplitude of the initial Wigner function through the time-dependent coefficients of the dynamical map. The resulting time-dependent Wigner function is given by
\begin{widetext}
\begin{align}
W_{|n\rangle}(x,p,t)&=\frac{(-1)^{n}}{\pi}\frac{1-g_{+}}{1+g_{+}}\exp\left(-R\frac{1-g_{+}}{1+g_{+}}\right)\left(\frac{e^{g_{0}}\Lambda}{1+g_{+}}\right)^{n}L_{n}\left[\frac{2 R }{(1+g_{+})\Lambda}\right],\label{Wigner_Fock_SU11}
\end{align}
\end{widetext}
with $R=x^{2}+p^{2}$, $\Lambda=1-e^{-g_{0}}g_{-}(1+g_{+})$, the coefficients $g_{0},g_{+},g_{-}$ obtained from Eq.~\eqref{g0}-\eqref{g_-} and the Laguerre polynomial $L_{n}(x)=\sum_{k=0}^{n}\binom{n}{k}\frac{(-x)^{k}}{k!}$. We verify that Eq.~\eqref{Wigner_Fock_SU11}, obtained from the exact propagator, agrees with the expression obtained from a phase-space analysis in Appendix~\ref{verification_Wigner_function}. An interesting limit is the pure--loss limit $(\gamma_{+}=0)$. In this situation
\begin{align}
g_{+}&=0,& g_{-}&=1-\upsilon,& g_{0}&=\ln(\upsilon),\label{pure_loss_limit}
\end{align}
with $\upsilon=e^{-\Gamma_{-}(t,0)}$. The Wigner function simplifies to 
\begin{align}
W_{|n\rangle}(x,p,t)&=\frac{(-1)^{n}}{\pi}\exp(-R)(2\upsilon-1)^{n} L_{n}\left(\frac{2\upsilon R}{2\upsilon-1}\right). \label{pure_loss_Wigner}
\end{align}
There is a singularity at $\upsilon=\frac{1}{2}$; however, it is removable by taking the limit $\upsilon\to \frac{1}{2}$, which gives
\begin{align}
{\rm lim}_{\upsilon\to \frac{1}{2}}\;W_{|n\rangle}(x,p,t)&=\frac{1}{\pi}e^{-R}\frac{R^{n}}{n!}.
\end{align}
To illustrate the deformation of the Wigner function under
nonadiabatic driving and dissipation, we use the master equation introduced in detail in Sec.~\ref{exp_setup}. Fig.~\ref{fig:Wigner_function_evolution} summarizes the driven protocols and resulting Wigner function dynamics. Panel (a) shows the compression (in red) and expansion protocols (in blue), while panel (b) shows the corresponding gain and loss rates (solid and dashed lines, respectively). Under compression, $\gamma_{+}$ decreases and $\gamma_{-}$ increases. By contrast, under expansion, $\gamma_{+}$ increases and $\gamma_{-}$ decreases. Panel (c) displays the Bose-Einstein occupation number Eq.~\eqref{occupation_number}. Panels (d)-(f) and (g)-(i) display the evolution of the Wigner function for a compression and expansion, respectively. Starting from an initial Fock state $|n\rangle$, the dissipative dynamics progressively suppresses the negative region of the Wigner function for both compression and expansion, reflecting the loss of nonclassicality induced by the thermal environment.

\subsection{Encoded-subspace sensitivity of a binomial bosonic code to driven gain--loss dynamics \label{bosonic_code}}

The driven harmonic oscillator also provides a natural setting in which to investigate the effect of time-dependent dissipation on bosonic quantum information. In bosonic quantum-error-correcting architectures, logical information is encoded directly in the Hilbert space of an oscillator, whose parameters may themselves be dynamically controlled. In particular, a time-dependent modulation of the oscillator frequency provides a natural example of a control protocol. As seen in Sec.~\ref{exp_setup}, such a modulation also modifies the dissipative dynamics experienced by the encoded state by rendering the photon-loss and photon-gain rates time dependent. It is therefore natural to ask how frequency modulation affects the survival probability of an encoded logical subspace. Here, we focus on this effect and quantify the resulting leakage from a fixed bosonic code space. The exact propagator derived above makes this question analytically accessible for arbitrary time-dependent gain and loss rates.
As a concrete example, we consider the binomial bosonic code \cite{Gottesman2001,Michael2016,Wang2026}, whose code words are designed to protect against photon loss and gain. Here, however, we restrict our analysis to leakage from the code space and do not consider recovery operations. We focus on the simplest binomial encoding satisfying the Knill--Laflamme conditions \cite{Nielsen_Chuang2010,Steane1996,Bennett1996,Laflamme1996} for the single-photon error set $\{\mathbbm{1},a,a^{\dagger}\}$ \cite{Michael2016}, as detailed in Appendix~\ref{app:binomial_correlations}
\begin{align}
|0_{L}\rangle
&=
\frac{|0\rangle+|6\rangle}{\sqrt{2}},
&
|1_{L}\rangle
&=
|3\rangle .
\label{eq:binomial_codewords}
\end{align}
The code-space projector is given by
\begin{align}
P_{C}
&=
|0_{L}\rangle\langle0_{L}|
+
|1_{L}\rangle\langle1_{L}|.
\label{eq:binomial_projector}
\end{align}
To characterize the action of the gain--loss dynamics on the code space, one can define the code-space survival probability 
\begin{align}
p_{C}(t)
&=
\operatorname{Tr}
\left[
P_{C}\rho(t)
\right].
\label{eq:code_survival_probability}
\end{align}
It gives the probability that the evolved oscillator state remains
supported on the encoded qubit subspace.

To evaluate this quantity, we introduce the Fock-space transition kernel $T_{rq|mn}$
of the dynamical map through
\begin{align}
V(t,s)
\left(
|m\rangle\langle n|
\right)
&=
\sum_{r,q=0}^{\infty}
T_{rq|mn}(t,s)
|r\rangle\langle q|.
\label{eq:transition_kernel_definition_main}
\end{align}
\begin{figure}
    \centering
    \includegraphics[width=\linewidth]{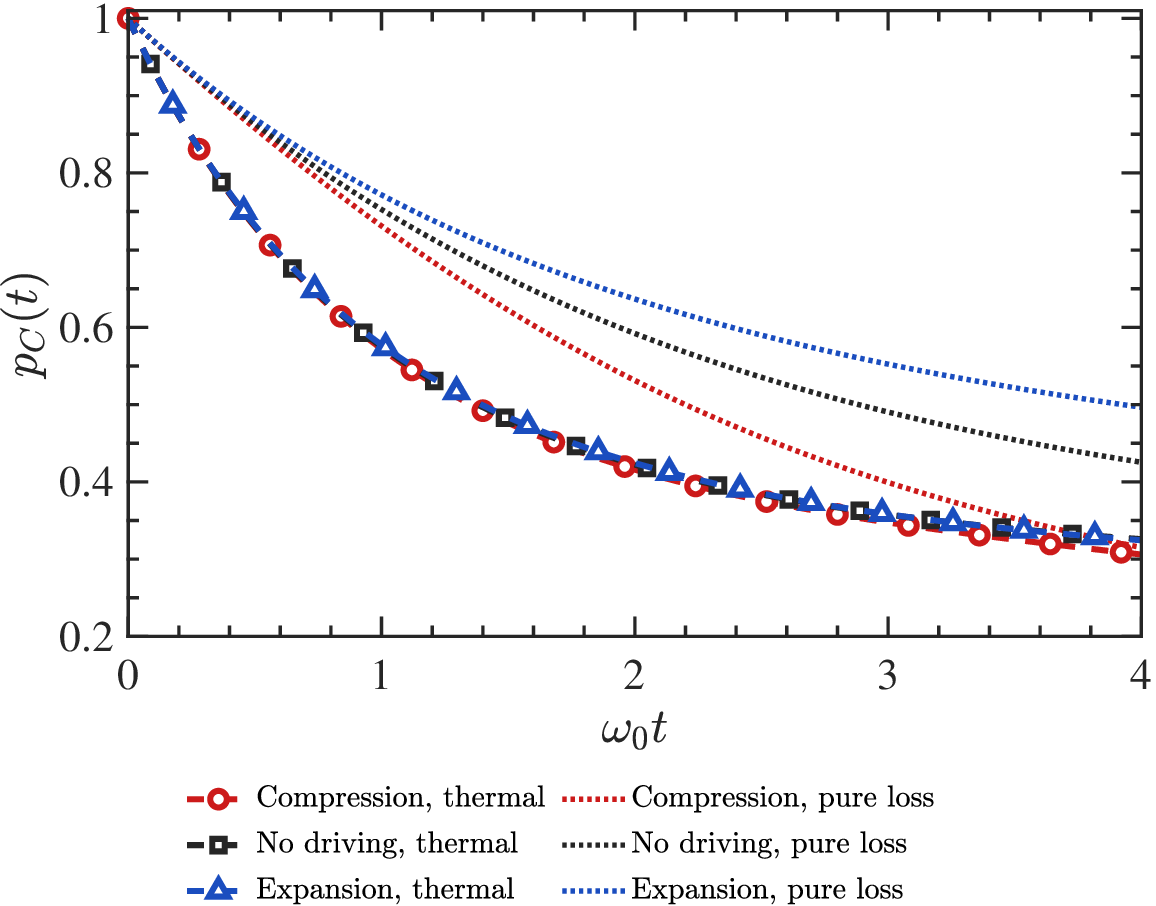}
    \caption{
    {\bf Code-space survival probability $p_C(t)$ for the binomial bosonic
    code under compression ($\xi=+0.20$), no driving ($\xi=0$), and
    expansion ($\xi=-0.20$)}. Dashed curves with markers show the finite-temperature
    gain--loss dynamics, while dotted curves show the corresponding
    pure--loss limits. Circles, squares, and triangles distinguish the
    compression, undriven, and expansion protocols, respectively.
    We use the same parameters as in Fig.~\ref{fig:Wigner_function_evolution}.
    }
\label{fig:code_survival_probability}
\end{figure}
Using the factorized propagator derived above, its matrix elements are
\begin{align}
T_{rq|mn}(t,s)
&=
\mathcal{N}(t,s)
\sum_{p=0}^{\min(m,n)}
\sum_{j=0}^{\infty}
e^{\frac{1}{2}(m+n-2p+1)g_{0}(t,s)}
\nonumber\\
&\quad\times
[g_{-}(t,s)]^{p}
[g_{+}(t,s)]^{j}
\delta_{r,m-p+j}
\delta_{q,n-p+j}
\nonumber\\
&\quad\times
\sqrt{
\binom{m}{p}
\binom{n}{p}
\binom{m-p+j}{m-p}
\binom{n-p+j}{n-p}
}.
\label{eq:transition_kernel}
\end{align}
%
%\vspace{-1cm}
%
We verify in Appendix~\ref{verification_birth_death} that the diagonal Fock state evolution agrees with that obtained independently from a birth--death master equation. For a general initial density matrix of elements $\rho(0)=\sum_{m,n}\rho_{mn}(0)|m\rangle\langle n|$, the survival probability can therefore be written as
\begin{align}
p_{C}(t)
&=
\sum_{r,q,m,n=0}^{\infty}
(P_{C})_{qr}
T_{rq|mn}(t,0)
\rho_{mn}(0), \label{survival_probability}
\end{align}
where $(P_{C})_{qr}=\langle q|P_{C}|r\rangle$. We provide the details of the calculations in Appendix~\ref{app:binomial_correlations}.

To further characterize the action of the channel without selecting a preferred
logical direction, we take the initial state to be maximally mixed within
the code space
\begin{align}
\rho(0)=\frac{P_{C}}{2}.
\label{eq:binomial_initial_state}
\end{align}
Introducing the compact notations
\begin{align}
\zeta&=e^{g_{0}/2},& \mu&=g_{-}, & \vartheta&=g_{+},
\end{align}
we obtain
\begin{align}
p_{C}(t)&=\frac{\mathcal{N}\zeta}{8}\Big[(1+\mu^{3})^{2}(1+\vartheta^{3})^{2}+36 \zeta^{2}\mu^{2}\vartheta^{2}(1+\mu^{3})(1+\vartheta^{3})\nonumber\\
&+9 \zeta^{4}\mu \vartheta (2+5\mu^{3})(2+5\vartheta^{3})+225 \zeta^{8}\mu^{2}\vartheta^{2}\nonumber\\
&+\zeta^{6}(6+40 \mu^{3}+40 \vartheta^{3}+400 \vartheta^{3}\mu^{3})+36 \zeta^{10} \mu \vartheta+\zeta^{12}\Big].\label{analytical_pC}
\end{align}
In the pure--loss limit given by Eq.~\eqref{pure_loss_limit},
$\zeta=\sqrt{\upsilon}$, $\mu=1-\upsilon$, $\vartheta=0$, and
$\mathcal{N}=\upsilon^{-1/2}$, yielding
\begin{align}
p_{C}(\upsilon)&=\frac{1}{8}\Big(4-12 \upsilon+21 \upsilon^{2}+24 \upsilon^{3}-105 \upsilon^{4}+114\upsilon^{5}-38 \upsilon^{6}\Big).
\end{align}

Figure~\ref{fig:code_survival_probability} illustrates how the driven
gain--loss dynamics modifies leakage from the logical subspace. We compare the pure--loss dynamics (in dotted lines) with finite-temperature gain--loss dynamics (colored markers). For pure--loss, the effect of the driving is significant, with the survival probability decaying at a slower rate for the expansion and at a faster rate for the compression relative to the undriven regime. In the thermal case, gain and loss act simultaneously. As shown in Fig.~\ref{fig:Wigner_function_evolution} (b), for a compression, $\gamma_{+}$ (gain) decreases and $\gamma_{-}$ (loss) increases. For an expansion, it is the opposite. Therefore, compression increases the loss rate, whereas expansion reduces it. This trend is consistent with Figure~\ref{fig:code_survival_probability}, with a survival probability slightly lower for the compression and very close to the undriven case for the expansion in the thermal regime. Hence, the binomial code appears to be weakly sensitive to the rate modulation in a thermal bath under these driving conditions. 
\section{Conclusion \label{conclusion}}
In this work, we derived explicit closed-form factorizations of chronological and antichronological propagators generated by two time-dependent operators whose commutator is a simultaneous eigenoperator of their adjoint actions. This closure reduces the time-ordered evolution to a finite product of ordinary exponentials and determines the factorization coefficients directly, without solving the coupled nonlinear differential equations that arise in the Wei--Norman construction. We applied this framework to the bosonic gain--loss Lindblad equation, whose dissipative generators satisfy the required algebraic closure, and further displayed an underlying $su(1,1)$ structure. The $su(1,1)$ structure is useful for computing the action of the propagator on an initial generalized Fock state. However, the algebra generated by the gain--loss Lindbladian is intrinsically two-dimensional, which explains the explicit solvability of the dynamics for arbitrary driving. This yields an exact dynamical map for time-dependent gain and loss rates and an arbitrary initial state. We used the resulting propagator to obtain explicit expressions for general Fock-state evolution and the associated Wigner function, and to derive the code-space survival probability of a binomial bosonic code under driven gain--loss dynamics. In particular, the binomial-code example shows that time-dependent rates can modify leakage from the encoded subspace in the pure--loss regime, but that the driving weakly affects the survival probability in a thermal bath. From a mathematical point of view, a natural next step is to determine how broadly the eigencommutator closure identified here extends beyond the gain--loss problem. More generally, identifying algebraic conditions under which the corresponding Wei--Norman equations are explicitly integrable could provide a systematic route to closed-form propagators for wider classes of time-dependent open-system generators, including dissipative models with nonlinear structure \cite{McDonald2023}. Another promising direction is to investigate state evolution in the presence of time-dependent negative rates.

\begin{acknowledgments}
The author thanks Stefano Scopa for useful comments, and Adolfo del Campo, Aurélia Chenu and Paola Ruggiero for mentorship. The author acknowledges support from the UK Engineering and Physical Sciences Research Council (EPSRC), grant number EP/Y015363/1. The author also acknowledges the University of Luxembourg for funding at the early stage of this work. The author used OpenAI ChatGPT (GPT-5.6 Sol) to assist with numerical-code development, equation checking, and limited language editing. All AI-assisted outputs were independently verified by the author, who takes full responsibility for the content of the manuscript.
\end{acknowledgments}

\bibliography{DOTUR_v2}

\begin{thebibliography}{67}%
\makeatletter
\providecommand \@ifxundefined [1]{%
 \@ifx{#1\undefined}
}%
\providecommand \@ifnum [1]{%
 \ifnum #1\expandafter \@firstoftwo
 \else \expandafter \@secondoftwo
 \fi
}%
\providecommand \@ifx [1]{%
 \ifx #1\expandafter \@firstoftwo
 \else \expandafter \@secondoftwo
 \fi
}%
\providecommand \natexlab [1]{#1}%
\providecommand \enquote  [1]{``#1''}%
\providecommand \bibnamefont  [1]{#1}%
\providecommand \bibfnamefont [1]{#1}%
\providecommand \citenamefont [1]{#1}%
\providecommand \href@noop [0]{\@secondoftwo}%
\providecommand \href [0]{\begingroup \@sanitize@url \@href}%
\providecommand \@href[1]{\@@startlink{#1}\@@href}%
\providecommand \@@href[1]{\endgroup#1\@@endlink}%
\providecommand \@sanitize@url [0]{\catcode `\\12\catcode `\$12\catcode
  `\&12\catcode `\#12\catcode `\^12\catcode `\_12\catcode `\%12\relax}%
\providecommand \@@startlink[1]{}%
\providecommand \@@endlink[0]{}%
\providecommand \url  [0]{\begingroup\@sanitize@url \@url }%
\providecommand \@url [1]{\endgroup\@href {#1}{\urlprefix }}%
\providecommand \urlprefix  [0]{URL }%
\providecommand \Eprint [0]{\href }%
\providecommand \doibase [0]{http://dx.doi.org/}%
\providecommand \selectlanguage [0]{\@gobble}%
\providecommand \bibinfo  [0]{\@secondoftwo}%
\providecommand \bibfield  [0]{\@secondoftwo}%
\providecommand \translation [1]{[#1]}%
\providecommand \BibitemOpen [0]{}%
\providecommand \bibitemStop [0]{}%
\providecommand \bibitemNoStop [0]{.\EOS\space}%
\providecommand \EOS [0]{\spacefactor3000\relax}%
\providecommand \BibitemShut  [1]{\csname bibitem#1\endcsname}%
\let\auto@bib@innerbib\@empty
%</preamble>
\bibitem [{\citenamefont {Verstraete}\ \emph {et~al.}(2009)\citenamefont
  {Verstraete}, \citenamefont {Wolf},\ and\ \citenamefont
  {Cirac}}]{Verstraete2009}%
  \BibitemOpen
  \bibfield  {author} {\bibinfo {author} {\bibfnamefont {F.}~\bibnamefont
  {Verstraete}}, \bibinfo {author} {\bibfnamefont {M.~M.}\ \bibnamefont
  {Wolf}}, \ and\ \bibinfo {author} {\bibfnamefont {J.~I.}\ \bibnamefont
  {Cirac}},\ }\bibfield  {title} {\enquote {\bibinfo {title} {Quantum
  computation and quantum-state engineering driven by dissipation},}\ }\href
  {\doibase 10.1038/nphys1342} {\bibfield  {journal} {\bibinfo  {journal} {Nat.
  Phys.}\ }\textbf {\bibinfo {volume} {5}},\ \bibinfo {pages} {633--636}
  (\bibinfo {year} {2009})}\BibitemShut {NoStop}%
\bibitem [{\citenamefont {Diehl}\ \emph {et~al.}(2008)\citenamefont {Diehl},
  \citenamefont {Micheli}, \citenamefont {Kantian}, \citenamefont {Kraus},
  \citenamefont {B{\"u}chler},\ and\ \citenamefont {Zoller}}]{Diehl2008}%
  \BibitemOpen
  \bibfield  {author} {\bibinfo {author} {\bibfnamefont {S.}~\bibnamefont
  {Diehl}}, \bibinfo {author} {\bibfnamefont {A.}~\bibnamefont {Micheli}},
  \bibinfo {author} {\bibfnamefont {A.}~\bibnamefont {Kantian}}, \bibinfo
  {author} {\bibfnamefont {B.}~\bibnamefont {Kraus}}, \bibinfo {author}
  {\bibfnamefont {H.~P.}\ \bibnamefont {B{\"u}chler}}, \ and\ \bibinfo {author}
  {\bibfnamefont {P.}~\bibnamefont {Zoller}},\ }\bibfield  {title} {\enquote
  {\bibinfo {title} {Quantum states and phases in driven open quantum systems
  with cold atoms},}\ }\href {\doibase 10.1038/nphys1073} {\bibfield  {journal}
  {\bibinfo  {journal} {Nature Phys.}\ }\textbf {\bibinfo {volume} {4}},\
  \bibinfo {pages} {878--883} (\bibinfo {year} {2008})}\BibitemShut {NoStop}%
\bibitem [{\citenamefont {Barreiro}\ \emph {et~al.}(2011)\citenamefont
  {Barreiro}, \citenamefont {M{\"u}ller}, \citenamefont {Schindler},
  \citenamefont {Nigg}, \citenamefont {Monz}, \citenamefont {Chwalla},
  \citenamefont {Hennrich}, \citenamefont {Roos}, \citenamefont {Zoller},\ and\
  \citenamefont {Blatt}}]{Barreiro2011}%
  \BibitemOpen
  \bibfield  {author} {\bibinfo {author} {\bibfnamefont {J.~T.}\ \bibnamefont
  {Barreiro}}, \bibinfo {author} {\bibfnamefont {M.}~\bibnamefont
  {M{\"u}ller}}, \bibinfo {author} {\bibfnamefont {P.}~\bibnamefont
  {Schindler}}, \bibinfo {author} {\bibfnamefont {D.}~\bibnamefont {Nigg}},
  \bibinfo {author} {\bibfnamefont {T.}~\bibnamefont {Monz}}, \bibinfo {author}
  {\bibfnamefont {M.}~\bibnamefont {Chwalla}}, \bibinfo {author} {\bibfnamefont
  {M.}~\bibnamefont {Hennrich}}, \bibinfo {author} {\bibfnamefont {C.~F.}\
  \bibnamefont {Roos}}, \bibinfo {author} {\bibfnamefont {P.}~\bibnamefont
  {Zoller}}, \ and\ \bibinfo {author} {\bibfnamefont {R.}~\bibnamefont
  {Blatt}},\ }\bibfield  {title} {\enquote {\bibinfo {title} {An open-system
  quantum simulator with trapped ions},}\ }\href {\doibase 10.1038/nature09801}
  {\bibfield  {journal} {\bibinfo  {journal} {Nature}\ }\textbf {\bibinfo
  {volume} {470}},\ \bibinfo {pages} {486--491} (\bibinfo {year}
  {2011})}\BibitemShut {NoStop}%
\bibitem [{\citenamefont {Shankar}\ \emph {et~al.}(2013)\citenamefont
  {Shankar}, \citenamefont {Hatridge}, \citenamefont {Leghtas}, \citenamefont
  {Sliwa}, \citenamefont {Narla}, \citenamefont {Vool}, \citenamefont {Girvin},
  \citenamefont {Frunzio}, \citenamefont {Mirrahimi},\ and\ \citenamefont
  {Devoret}}]{Shankar2013}%
  \BibitemOpen
  \bibfield  {author} {\bibinfo {author} {\bibfnamefont {S.}~\bibnamefont
  {Shankar}}, \bibinfo {author} {\bibfnamefont {M.}~\bibnamefont {Hatridge}},
  \bibinfo {author} {\bibfnamefont {Z.}~\bibnamefont {Leghtas}}, \bibinfo
  {author} {\bibfnamefont {K.~M.}\ \bibnamefont {Sliwa}}, \bibinfo {author}
  {\bibfnamefont {A.}~\bibnamefont {Narla}}, \bibinfo {author} {\bibfnamefont
  {U.}~\bibnamefont {Vool}}, \bibinfo {author} {\bibfnamefont {S.~M.}\
  \bibnamefont {Girvin}}, \bibinfo {author} {\bibfnamefont {L.}~\bibnamefont
  {Frunzio}}, \bibinfo {author} {\bibfnamefont {M.}~\bibnamefont {Mirrahimi}},
  \ and\ \bibinfo {author} {\bibfnamefont {M.~H.}\ \bibnamefont {Devoret}},\
  }\bibfield  {title} {\enquote {\bibinfo {title} {Autonomously stabilized
  entanglement between two superconducting quantum bits},}\ }\href {\doibase
  10.1038/nature12802} {\bibfield  {journal} {\bibinfo  {journal} {Nature}\
  }\textbf {\bibinfo {volume} {504}},\ \bibinfo {pages} {419--422} (\bibinfo
  {year} {2013})}\BibitemShut {NoStop}%
\bibitem [{\citenamefont {Vacanti}\ \emph {et~al.}(2014)\citenamefont
  {Vacanti}, \citenamefont {Fazio}, \citenamefont {Montangero}, \citenamefont
  {Palma}, \citenamefont {Paternostro},\ and\ \citenamefont
  {Vedral}}]{Vacanti2014}%
  \BibitemOpen
  \bibfield  {author} {\bibinfo {author} {\bibfnamefont {G}~\bibnamefont
  {Vacanti}}, \bibinfo {author} {\bibfnamefont {R}~\bibnamefont {Fazio}},
  \bibinfo {author} {\bibfnamefont {S}~\bibnamefont {Montangero}}, \bibinfo
  {author} {\bibfnamefont {G~M}\ \bibnamefont {Palma}}, \bibinfo {author}
  {\bibfnamefont {M}~\bibnamefont {Paternostro}}, \ and\ \bibinfo {author}
  {\bibfnamefont {V}~\bibnamefont {Vedral}},\ }\bibfield  {title} {\enquote
  {\bibinfo {title} {Transitionless quantum driving in open quantum systems},}\
  }\href {\doibase 10.1088/1367-2630/16/5/053017} {\bibfield  {journal}
  {\bibinfo  {journal} {New J. Phys.}\ }\textbf {\bibinfo {volume} {16}},\
  \bibinfo {pages} {053017} (\bibinfo {year} {2014})}\BibitemShut {NoStop}%
\bibitem [{\citenamefont {Alipour}\ \emph {et~al.}(2020)\citenamefont
  {Alipour}, \citenamefont {Chenu}, \citenamefont {Rezakhani},\ and\
  \citenamefont {del Campo}}]{Alipour2020}%
  \BibitemOpen
  \bibfield  {author} {\bibinfo {author} {\bibfnamefont {Sahar}\ \bibnamefont
  {Alipour}}, \bibinfo {author} {\bibfnamefont {Aurelia}\ \bibnamefont
  {Chenu}}, \bibinfo {author} {\bibfnamefont {Ali~T.}\ \bibnamefont
  {Rezakhani}}, \ and\ \bibinfo {author} {\bibfnamefont {Adolfo}\ \bibnamefont
  {del Campo}},\ }\bibfield  {title} {\enquote {\bibinfo {title} {Shortcuts to
  {A}diabaticity in {D}riven {O}pen {Q}uantum {S}ystems: {B}alanced {G}ain and
  {L}oss and {N}on-{M}arkovian {E}volution},}\ }\href {\doibase
  10.22331/q-2020-09-28-336} {\bibfield  {journal} {\bibinfo  {journal}
  {{Quantum}}\ }\textbf {\bibinfo {volume} {4}},\ \bibinfo {pages} {336}
  (\bibinfo {year} {2020})}\BibitemShut {NoStop}%
\bibitem [{\citenamefont {Sieberer}\ \emph {et~al.}(2016)\citenamefont
  {Sieberer}, \citenamefont {Buchhold},\ and\ \citenamefont
  {Diehl}}]{Sieberer2016}%
  \BibitemOpen
  \bibfield  {author} {\bibinfo {author} {\bibfnamefont {L.~M}\ \bibnamefont
  {Sieberer}}, \bibinfo {author} {\bibfnamefont {M.}~\bibnamefont {Buchhold}},
  \ and\ \bibinfo {author} {\bibfnamefont {S.}~\bibnamefont {Diehl}},\
  }\bibfield  {title} {\enquote {\bibinfo {title} {Keldysh field theory for
  driven open quantum systems},}\ }\href {\doibase
  10.1088/0034-4885/79/9/096001} {\bibfield  {journal} {\bibinfo  {journal}
  {Rep. Prog. Phys.}\ }\textbf {\bibinfo {volume} {79}},\ \bibinfo {pages}
  {096001} (\bibinfo {year} {2016})}\BibitemShut {NoStop}%
\bibitem [{\citenamefont {Kalthoff}\ \emph {et~al.}(2022)\citenamefont
  {Kalthoff}, \citenamefont {Kennes}, \citenamefont {Millis},\ and\
  \citenamefont {Sentef}}]{Kalthoff2022}%
  \BibitemOpen
  \bibfield  {author} {\bibinfo {author} {\bibfnamefont {M.~H.}\ \bibnamefont
  {Kalthoff}}, \bibinfo {author} {\bibfnamefont {D.~M.}\ \bibnamefont
  {Kennes}}, \bibinfo {author} {\bibfnamefont {A.~J.}\ \bibnamefont {Millis}},
  \ and\ \bibinfo {author} {\bibfnamefont {M.~A.}\ \bibnamefont {Sentef}},\
  }\bibfield  {title} {\enquote {\bibinfo {title} {Nonequilibrium phase
  transition in a driven-dissipative quantum antiferromagnet},}\ }\href
  {\doibase 10.1103/PhysRevResearch.4.023115} {\bibfield  {journal} {\bibinfo
  {journal} {Phys. Rev. Res.}\ }\textbf {\bibinfo {volume} {4}},\ \bibinfo
  {pages} {023115} (\bibinfo {year} {2022})}\BibitemShut {NoStop}%
\bibitem [{\citenamefont {Rakovszky}\ \emph {et~al.}(2024)\citenamefont
  {Rakovszky}, \citenamefont {Gopalakrishnan},\ and\ \citenamefont {von
  Keyserlingk}}]{Rakovszky2024}%
  \BibitemOpen
  \bibfield  {author} {\bibinfo {author} {\bibfnamefont {T.}~\bibnamefont
  {Rakovszky}}, \bibinfo {author} {\bibfnamefont {S.}~\bibnamefont
  {Gopalakrishnan}}, \ and\ \bibinfo {author} {\bibfnamefont {C.}~\bibnamefont
  {von Keyserlingk}},\ }\bibfield  {title} {\enquote {\bibinfo {title}
  {Defining stable phases of open quantum systems},}\ }\href {\doibase
  10.1103/PhysRevX.14.041031} {\bibfield  {journal} {\bibinfo  {journal} {Phys.
  Rev. X}\ }\textbf {\bibinfo {volume} {14}},\ \bibinfo {pages} {041031}
  (\bibinfo {year} {2024})}\BibitemShut {NoStop}%
\bibitem [{\citenamefont {Kosloff}(2013)}]{Kosloff2013}%
  \BibitemOpen
  \bibfield  {author} {\bibinfo {author} {\bibfnamefont {R.}~\bibnamefont
  {Kosloff}},\ }\bibfield  {title} {\enquote {\bibinfo {title} {Quantum
  thermodynamics: A dynamical viewpoint},}\ }\href {\doibase 10.3390/e15062100}
  {\bibfield  {journal} {\bibinfo  {journal} {Entropy}\ }\textbf {\bibinfo
  {volume} {15}},\ \bibinfo {pages} {2100--2128} (\bibinfo {year}
  {2013})}\BibitemShut {NoStop}%
\bibitem [{\citenamefont {Benenti}\ \emph {et~al.}(2017)\citenamefont
  {Benenti}, \citenamefont {Casati}, \citenamefont {Saito},\ and\ \citenamefont
  {Whitney}}]{Benenti2017}%
  \BibitemOpen
  \bibfield  {author} {\bibinfo {author} {\bibfnamefont {G.}~\bibnamefont
  {Benenti}}, \bibinfo {author} {\bibfnamefont {G.}~\bibnamefont {Casati}},
  \bibinfo {author} {\bibfnamefont {K.}~\bibnamefont {Saito}}, \ and\ \bibinfo
  {author} {\bibfnamefont {R.~S.}\ \bibnamefont {Whitney}},\ }\bibfield
  {title} {\enquote {\bibinfo {title} {Fundamental aspects of steady-state
  conversion of heat to work at the nanoscale},}\ }\href {\doibase
  https://doi.org/10.1016/j.physrep.2017.05.008} {\bibfield  {journal}
  {\bibinfo  {journal} {Phys. Rep.}\ }\textbf {\bibinfo {volume} {694}},\
  \bibinfo {pages} {1--124} (\bibinfo {year} {2017})}\BibitemShut {NoStop}%
\bibitem [{\citenamefont {Myers}\ \emph {et~al.}(2022)\citenamefont {Myers},
  \citenamefont {Abah},\ and\ \citenamefont {Deffner}}]{Myers2022}%
  \BibitemOpen
  \bibfield  {author} {\bibinfo {author} {\bibfnamefont {N.~M.}\ \bibnamefont
  {Myers}}, \bibinfo {author} {\bibfnamefont {O.}~\bibnamefont {Abah}}, \ and\
  \bibinfo {author} {\bibfnamefont {S.}~\bibnamefont {Deffner}},\ }\bibfield
  {title} {\enquote {\bibinfo {title} {Quantum thermodynamic devices: From
  theoretical proposals to experimental reality},}\ }\href {\doibase
  10.1116/5.0083192} {\bibfield  {journal} {\bibinfo  {journal} {AVS Quantum
  Sci.}\ }\textbf {\bibinfo {volume} {4}},\ \bibinfo {pages} {027101} (\bibinfo
  {year} {2022})}\BibitemShut {NoStop}%
\bibitem [{\citenamefont {Landi}\ \emph {et~al.}(2022)\citenamefont {Landi},
  \citenamefont {Poletti},\ and\ \citenamefont {Schaller}}]{Landi2022}%
  \BibitemOpen
  \bibfield  {author} {\bibinfo {author} {\bibfnamefont {G.~T.}\ \bibnamefont
  {Landi}}, \bibinfo {author} {\bibfnamefont {D.}~\bibnamefont {Poletti}}, \
  and\ \bibinfo {author} {\bibfnamefont {G.}~\bibnamefont {Schaller}},\
  }\bibfield  {title} {\enquote {\bibinfo {title} {Nonequilibrium
  boundary-driven quantum systems: Models, methods, and properties},}\ }\href
  {\doibase 10.1103/RevModPhys.94.045006} {\bibfield  {journal} {\bibinfo
  {journal} {Rev. Mod. Phys.}\ }\textbf {\bibinfo {volume} {94}},\ \bibinfo
  {pages} {045006} (\bibinfo {year} {2022})}\BibitemShut {NoStop}%
\bibitem [{\citenamefont {Popkov}\ \emph {et~al.}(2020)\citenamefont {Popkov},
  \citenamefont {Prosen},\ and\ \citenamefont {Zadnik}}]{Popkov2020}%
  \BibitemOpen
  \bibfield  {author} {\bibinfo {author} {\bibfnamefont {V.}~\bibnamefont
  {Popkov}}, \bibinfo {author} {\bibfnamefont {T.}~\bibnamefont {Prosen}}, \
  and\ \bibinfo {author} {\bibfnamefont {L.}~\bibnamefont {Zadnik}},\
  }\bibfield  {title} {\enquote {\bibinfo {title} {Exact nonequilibrium steady
  state of open $xxz/xyz$ spin-$1/2$ chain with dirichlet boundary
  conditions},}\ }\href {\doibase 10.1103/PhysRevLett.124.160403} {\bibfield
  {journal} {\bibinfo  {journal} {Phys. Rev. Lett.}\ }\textbf {\bibinfo
  {volume} {124}},\ \bibinfo {pages} {160403} (\bibinfo {year}
  {2020})}\BibitemShut {NoStop}%
\bibitem [{\citenamefont {Haddadfarshi}\ \emph {et~al.}(2015)\citenamefont
  {Haddadfarshi}, \citenamefont {Cui},\ and\ \citenamefont
  {Mintert}}]{Haddadfarshi2015}%
  \BibitemOpen
  \bibfield  {author} {\bibinfo {author} {\bibfnamefont {F.}~\bibnamefont
  {Haddadfarshi}}, \bibinfo {author} {\bibfnamefont {J.}~\bibnamefont {Cui}}, \
  and\ \bibinfo {author} {\bibfnamefont {F.}~\bibnamefont {Mintert}},\
  }\bibfield  {title} {\enquote {\bibinfo {title} {Completely positive
  approximate solutions of driven open quantum systems},}\ }\href {\doibase
  10.1103/PhysRevLett.114.130402} {\bibfield  {journal} {\bibinfo  {journal}
  {Phys. Rev. Lett.}\ }\textbf {\bibinfo {volume} {114}},\ \bibinfo {pages}
  {130402} (\bibinfo {year} {2015})}\BibitemShut {NoStop}%
\bibitem [{\citenamefont {Schnell}\ \emph {et~al.}(2020)\citenamefont
  {Schnell}, \citenamefont {Eckardt},\ and\ \citenamefont
  {Denisov}}]{Schnell2020}%
  \BibitemOpen
  \bibfield  {author} {\bibinfo {author} {\bibfnamefont {A.}~\bibnamefont
  {Schnell}}, \bibinfo {author} {\bibfnamefont {A.}~\bibnamefont {Eckardt}}, \
  and\ \bibinfo {author} {\bibfnamefont {S.}~\bibnamefont {Denisov}},\
  }\bibfield  {title} {\enquote {\bibinfo {title} {Is there a floquet
  lindbladian?}}\ }\href {\doibase 10.1103/PhysRevB.101.100301} {\bibfield
  {journal} {\bibinfo  {journal} {Phys. Rev. B}\ }\textbf {\bibinfo {volume}
  {101}},\ \bibinfo {pages} {100301(R)} (\bibinfo {year} {2020})}\BibitemShut
  {NoStop}%
\bibitem [{\citenamefont {Dupays}\ \emph {et~al.}(2020)\citenamefont {Dupays},
  \citenamefont {Egusquiza}, \citenamefont {del Campo},\ and\ \citenamefont
  {Chenu}}]{Dupays2020}%
  \BibitemOpen
  \bibfield  {author} {\bibinfo {author} {\bibfnamefont {L.}~\bibnamefont
  {Dupays}}, \bibinfo {author} {\bibfnamefont {I.~L.}\ \bibnamefont
  {Egusquiza}}, \bibinfo {author} {\bibfnamefont {A.}~\bibnamefont {del
  Campo}}, \ and\ \bibinfo {author} {\bibfnamefont {A.}~\bibnamefont {Chenu}},\
  }\bibfield  {title} {\enquote {\bibinfo {title} {Superadiabatic
  thermalization of a quantum oscillator by engineered dephasing},}\ }\href
  {\doibase 10.1103/PhysRevResearch.2.033178} {\bibfield  {journal} {\bibinfo
  {journal} {Phys. Rev. Res.}\ }\textbf {\bibinfo {volume} {2}},\ \bibinfo
  {pages} {033178} (\bibinfo {year} {2020})}\BibitemShut {NoStop}%
\bibitem [{\citenamefont {Dupays}\ and\ \citenamefont
  {Chenu}(2021)}]{Dupays2021}%
  \BibitemOpen
  \bibfield  {author} {\bibinfo {author} {\bibfnamefont {L.}~\bibnamefont
  {Dupays}}\ and\ \bibinfo {author} {\bibfnamefont {A.}~\bibnamefont {Chenu}},\
  }\bibfield  {title} {\enquote {\bibinfo {title} {Shortcuts to {S}queezed
  {T}hermal {S}tates},}\ }\href {\doibase 10.22331/q-2021-05-01-449} {\bibfield
   {journal} {\bibinfo  {journal} {{Quantum}}\ }\textbf {\bibinfo {volume}
  {5}},\ \bibinfo {pages} {449} (\bibinfo {year} {2021})}\BibitemShut {NoStop}%
\bibitem [{\citenamefont {Mori}(2023)}]{Mori2023}%
  \BibitemOpen
  \bibfield  {author} {\bibinfo {author} {\bibfnamefont {T.}~\bibnamefont
  {Mori}},\ }\bibfield  {title} {\enquote {\bibinfo {title} {Floquet states in
  open quantum systems},}\ }\href {\doibase
  https://doi.org/10.1146/annurev-conmatphys-040721-015537} {\bibfield
  {journal} {\bibinfo  {journal} {Annu. Rev. Condens. Matter Phys.}\ }\textbf
  {\bibinfo {volume} {14}},\ \bibinfo {pages} {35--56} (\bibinfo {year}
  {2023})}\BibitemShut {NoStop}%
\bibitem [{\citenamefont {Dann}(2025)}]{Dann2025}%
  \BibitemOpen
  \bibfield  {author} {\bibinfo {author} {\bibfnamefont {R.}~\bibnamefont
  {Dann}},\ }\bibfield  {title} {\enquote {\bibinfo {title} {Interplay between
  external driving, dissipation and collective effects in the {M}arkovian and
  non-{M}arkovian regimes},}\ }\href {\doibase 10.22331/q-2025-05-12-1740}
  {\bibfield  {journal} {\bibinfo  {journal} {{Quantum}}\ }\textbf {\bibinfo
  {volume} {9}},\ \bibinfo {pages} {1740} (\bibinfo {year} {2025})}\BibitemShut
  {NoStop}%
\bibitem [{\citenamefont {Keliri}\ and\ \citenamefont
  {Schirò}(2026)}]{Keliri2026}%
  \BibitemOpen
  \bibfield  {author} {\bibinfo {author} {\bibfnamefont {A.}~\bibnamefont
  {Keliri}}\ and\ \bibinfo {author} {\bibfnamefont {M.}~\bibnamefont
  {Schirò}},\ }\bibfield  {title} {\enquote {\bibinfo {title} {Sambe approach
  to floquet-lindblad open quantum systems},}\ }\href
  {https://arxiv.org/abs/2606.09727} {\  (\bibinfo {year} {2026})},\ \Eprint
  {http://arxiv.org/abs/2606.09727} {arXiv:2606.09727 [quant-ph]} \BibitemShut
  {NoStop}%
\bibitem [{\citenamefont {Kamleitner}\ and\ \citenamefont
  {Shnirman}(2011)}]{Kamleitner2011}%
  \BibitemOpen
  \bibfield  {author} {\bibinfo {author} {\bibfnamefont {I.}~\bibnamefont
  {Kamleitner}}\ and\ \bibinfo {author} {\bibfnamefont {A.}~\bibnamefont
  {Shnirman}},\ }\bibfield  {title} {\enquote {\bibinfo {title} {Time-dependent
  markovian master equation for adiabatic systems and its application to
  cooper-pair pumping},}\ }\href {\doibase 10.1103/PhysRevB.84.235140}
  {\bibfield  {journal} {\bibinfo  {journal} {Phys. Rev. B}\ }\textbf {\bibinfo
  {volume} {84}},\ \bibinfo {pages} {235140} (\bibinfo {year}
  {2011})}\BibitemShut {NoStop}%
\bibitem [{\citenamefont {Reimer}\ \emph {et~al.}(2018)\citenamefont {Reimer},
  \citenamefont {Pedersen}, \citenamefont {Tanger}, \citenamefont
  {Pletyukhov},\ and\ \citenamefont {Gritsev}}]{Reimer2018}%
  \BibitemOpen
  \bibfield  {author} {\bibinfo {author} {\bibfnamefont {V.}~\bibnamefont
  {Reimer}}, \bibinfo {author} {\bibfnamefont {K.~G.~L.}\ \bibnamefont
  {Pedersen}}, \bibinfo {author} {\bibfnamefont {N.}~\bibnamefont {Tanger}},
  \bibinfo {author} {\bibfnamefont {M.}~\bibnamefont {Pletyukhov}}, \ and\
  \bibinfo {author} {\bibfnamefont {V.}~\bibnamefont {Gritsev}},\ }\bibfield
  {title} {\enquote {\bibinfo {title} {Nonadiabatic effects in periodically
  driven dissipative open quantum systems},}\ }\href {\doibase
  10.1103/PhysRevA.97.043851} {\bibfield  {journal} {\bibinfo  {journal} {Phys.
  Rev. A}\ }\textbf {\bibinfo {volume} {97}},\ \bibinfo {pages} {043851}
  (\bibinfo {year} {2018})}\BibitemShut {NoStop}%
\bibitem [{\citenamefont {Dann}\ \emph {et~al.}(2019)\citenamefont {Dann},
  \citenamefont {Tobalina},\ and\ \citenamefont {Kosloff}}]{Dann2019}%
  \BibitemOpen
  \bibfield  {author} {\bibinfo {author} {\bibfnamefont {R.}~\bibnamefont
  {Dann}}, \bibinfo {author} {\bibfnamefont {A.}~\bibnamefont {Tobalina}}, \
  and\ \bibinfo {author} {\bibfnamefont {R.}~\bibnamefont {Kosloff}},\
  }\bibfield  {title} {\enquote {\bibinfo {title} {Shortcut to equilibration of
  an open quantum system},}\ }\href {\doibase 10.1103/PhysRevLett.122.250402}
  {\bibfield  {journal} {\bibinfo  {journal} {Phys. Rev. Lett.}\ }\textbf
  {\bibinfo {volume} {122}},\ \bibinfo {pages} {250402} (\bibinfo {year}
  {2019})}\BibitemShut {NoStop}%
\bibitem [{\citenamefont {Shavit}\ \emph {et~al.}(2019)\citenamefont {Shavit},
  \citenamefont {Horovitz},\ and\ \citenamefont {Goldstein}}]{Shavit2019}%
  \BibitemOpen
  \bibfield  {author} {\bibinfo {author} {\bibfnamefont {G.}~\bibnamefont
  {Shavit}}, \bibinfo {author} {\bibfnamefont {B.}~\bibnamefont {Horovitz}}, \
  and\ \bibinfo {author} {\bibfnamefont {M.}~\bibnamefont {Goldstein}},\
  }\bibfield  {title} {\enquote {\bibinfo {title} {Bridging between laboratory
  and rotating-frame master equations for open quantum systems},}\ }\href
  {\doibase 10.1103/PhysRevB.100.195436} {\bibfield  {journal} {\bibinfo
  {journal} {Phys. Rev. B}\ }\textbf {\bibinfo {volume} {100}},\ \bibinfo
  {pages} {195436} (\bibinfo {year} {2019})}\BibitemShut {NoStop}%
\bibitem [{\citenamefont {Wu}\ \emph {et~al.}(2022)\citenamefont {Wu},
  \citenamefont {Huang},\ and\ \citenamefont {Yi}}]{Wu2022}%
  \BibitemOpen
  \bibfield  {author} {\bibinfo {author} {\bibfnamefont {S.~L.}\ \bibnamefont
  {Wu}}, \bibinfo {author} {\bibfnamefont {X.~L.}\ \bibnamefont {Huang}}, \
  and\ \bibinfo {author} {\bibfnamefont {X.~X.}\ \bibnamefont {Yi}},\
  }\bibfield  {title} {\enquote {\bibinfo {title} {Driven markovian master
  equation based on the lewis-riesenfeld-invariant theory},}\ }\href {\doibase
  10.1103/PhysRevA.106.052217} {\bibfield  {journal} {\bibinfo  {journal}
  {Phys. Rev. A}\ }\textbf {\bibinfo {volume} {106}},\ \bibinfo {pages}
  {052217} (\bibinfo {year} {2022})}\BibitemShut {NoStop}%
\bibitem [{\citenamefont {Di~Meglio}\ \emph {et~al.}(2024)\citenamefont
  {Di~Meglio}, \citenamefont {Plenio},\ and\ \citenamefont
  {Huelga}}]{DiMeglio2024}%
  \BibitemOpen
  \bibfield  {author} {\bibinfo {author} {\bibfnamefont {G.}~\bibnamefont
  {Di~Meglio}}, \bibinfo {author} {\bibfnamefont {M.~B.}\ \bibnamefont
  {Plenio}}, \ and\ \bibinfo {author} {\bibfnamefont {S.~F.}\ \bibnamefont
  {Huelga}},\ }\bibfield  {title} {\enquote {\bibinfo {title} {Time dependent
  {M}arkovian master equation beyond the adiabatic limit},}\ }\href {\doibase
  10.22331/q-2024-11-21-1534} {\bibfield  {journal} {\bibinfo  {journal}
  {{Quantum}}\ }\textbf {\bibinfo {volume} {8}},\ \bibinfo {pages} {1534}
  (\bibinfo {year} {2024})}\BibitemShut {NoStop}%
\bibitem [{\citenamefont {Gorini}\ \emph {et~al.}(1976)\citenamefont {Gorini},
  \citenamefont {Kossakowski},\ and\ \citenamefont {Sudarshan}}]{Gorini1976}%
  \BibitemOpen
  \bibfield  {author} {\bibinfo {author} {\bibfnamefont {V.}~\bibnamefont
  {Gorini}}, \bibinfo {author} {\bibfnamefont {A.}~\bibnamefont {Kossakowski}},
  \ and\ \bibinfo {author} {\bibfnamefont {E.~C.~G.}\ \bibnamefont
  {Sudarshan}},\ }\bibfield  {title} {\enquote {\bibinfo {title} {Completely
  positive dynamical semigroups of n‐level systems},}\ }\href {\doibase
  10.1063/1.522979} {\bibfield  {journal} {\bibinfo  {journal} {J. Math.
  Phys.}\ }\textbf {\bibinfo {volume} {17}},\ \bibinfo {pages} {821--825}
  (\bibinfo {year} {1976})}\BibitemShut {NoStop}%
\bibitem [{\citenamefont {Lindblad}(1976)}]{Lindblad1976}%
  \BibitemOpen
  \bibfield  {author} {\bibinfo {author} {\bibfnamefont {G.}~\bibnamefont
  {Lindblad}},\ }\bibfield  {title} {\enquote {\bibinfo {title} {On the
  generators of quantum dynamical semigroups},}\ }\href {\doibase
  10.1007/BF01608499} {\bibfield  {journal} {\bibinfo  {journal} {Commun. Math.
  Phys.}\ }\textbf {\bibinfo {volume} {48}},\ \bibinfo {pages} {119--130}
  (\bibinfo {year} {1976})}\BibitemShut {NoStop}%
\bibitem [{\citenamefont {Stefanini}\ \emph {et~al.}(2026)\citenamefont
  {Stefanini}, \citenamefont {Ziolkowska}, \citenamefont {Budker},
  \citenamefont {Poschinger}, \citenamefont {Schmidt-Kaler}, \citenamefont
  {Browaeys}, \citenamefont {Imamoglu}, \citenamefont {Chang},\ and\
  \citenamefont {Marino}}]{Stefanini2026}%
  \BibitemOpen
  \bibfield  {author} {\bibinfo {author} {\bibfnamefont {M.}~\bibnamefont
  {Stefanini}}, \bibinfo {author} {\bibfnamefont {A.~A.}\ \bibnamefont
  {Ziolkowska}}, \bibinfo {author} {\bibfnamefont {D.}~\bibnamefont {Budker}},
  \bibinfo {author} {\bibfnamefont {U.}~\bibnamefont {Poschinger}}, \bibinfo
  {author} {\bibfnamefont {F.}~\bibnamefont {Schmidt-Kaler}}, \bibinfo {author}
  {\bibfnamefont {A.}~\bibnamefont {Browaeys}}, \bibinfo {author}
  {\bibfnamefont {A.}~\bibnamefont {Imamoglu}}, \bibinfo {author}
  {\bibfnamefont {D.}~\bibnamefont {Chang}}, \ and\ \bibinfo {author}
  {\bibfnamefont {J.}~\bibnamefont {Marino}},\ }\bibfield  {title} {\enquote
  {\bibinfo {title} {Is lindblad for me?}}\ }\href {\doibase
  10.21468/SciPostPhysLectNotes.129} {\bibfield  {journal} {\bibinfo  {journal}
  {SciPost Phys. Lect. Notes}\ ,\ \bibinfo {pages} {129}} (\bibinfo {year}
  {2026})}\BibitemShut {NoStop}%
\bibitem [{\citenamefont {Trotter}(1959)}]{Trotter1959}%
  \BibitemOpen
  \bibfield  {author} {\bibinfo {author} {\bibfnamefont {H.~F.}\ \bibnamefont
  {Trotter}},\ }\bibfield  {title} {\enquote {\bibinfo {title} {On the product
  of semi-groups of operators},}\ }\href {\doibase
  10.1090/s0002-9939-1959-0108732-6} {\bibfield  {journal} {\bibinfo  {journal}
  {Proc. Amer. Math. Soc.}\ }\textbf {\bibinfo {volume} {10}},\ \bibinfo
  {pages} {545–551} (\bibinfo {year} {1959})}\BibitemShut {NoStop}%
\bibitem [{\citenamefont {Suzuki}(1976)}]{Suzuki1976}%
  \BibitemOpen
  \bibfield  {author} {\bibinfo {author} {\bibfnamefont {M.}~\bibnamefont
  {Suzuki}},\ }\bibfield  {title} {\enquote {\bibinfo {title} {Generalized
  trotter's formula and systematic approximants of exponential operators and
  inner derivations with applications to many-body problems},}\ }\href
  {\doibase 10.1007/BF01609348} {\bibfield  {journal} {\bibinfo  {journal}
  {Commun. Math. Phys.}\ }\textbf {\bibinfo {volume} {51}},\ \bibinfo {pages}
  {183--190} (\bibinfo {year} {1976})}\BibitemShut {NoStop}%
\bibitem [{\citenamefont {Suzuki}(1990)}]{Suzuki1990}%
  \BibitemOpen
  \bibfield  {author} {\bibinfo {author} {\bibfnamefont {M.}~\bibnamefont
  {Suzuki}},\ }\bibfield  {title} {\enquote {\bibinfo {title} {Fractal
  decomposition of exponential operators with applications to many-body
  theories and monte carlo simulations},}\ }\href {\doibase
  https://doi.org/10.1016/0375-9601(90)90962-N} {\bibfield  {journal} {\bibinfo
   {journal} {Phys. Lett. A}\ }\textbf {\bibinfo {volume} {146}},\ \bibinfo
  {pages} {319--323} (\bibinfo {year} {1990})}\BibitemShut {NoStop}%
\bibitem [{\citenamefont {Huyghebaert}\ and\ \citenamefont
  {De~Raedt}(1990)}]{Huyghebaert1990}%
  \BibitemOpen
  \bibfield  {author} {\bibinfo {author} {\bibfnamefont {J.}~\bibnamefont
  {Huyghebaert}}\ and\ \bibinfo {author} {\bibfnamefont {H.}~\bibnamefont
  {De~Raedt}},\ }\bibfield  {title} {\enquote {\bibinfo {title} {Product
  formula methods for time-dependent schr{\"o}dinger problems},}\ }\href
  {\doibase 10.1088/0305-4470/23/24/019} {\bibfield  {journal} {\bibinfo
  {journal} {J. Phys. A: Math. Gen.}\ }\textbf {\bibinfo {volume} {23}},\
  \bibinfo {pages} {5777--5793} (\bibinfo {year} {1990})}\BibitemShut {NoStop}%
\bibitem [{\citenamefont {David}\ \emph {et~al.}(2025)\citenamefont {David},
  \citenamefont {Sinayskiy},\ and\ \citenamefont {Petruccione}}]{David2025}%
  \BibitemOpen
  \bibfield  {author} {\bibinfo {author} {\bibfnamefont {I.~J.}\ \bibnamefont
  {David}}, \bibinfo {author} {\bibfnamefont {I.}~\bibnamefont {Sinayskiy}}, \
  and\ \bibinfo {author} {\bibfnamefont {F.}~\bibnamefont {Petruccione}},\
  }\bibfield  {title} {\enquote {\bibinfo {title} {Faster quantum simulation of
  markovian open quantum systems via randomisation},}\ }\href
  {https://arxiv.org/abs/2408.11683} {\  (\bibinfo {year} {2025})},\ \Eprint
  {http://arxiv.org/abs/2408.11683} {arXiv:2408.11683 [quant-ph]} \BibitemShut
  {NoStop}%
\bibitem [{\citenamefont {Pillay}\ \emph {et~al.}(2026)\citenamefont {Pillay},
  \citenamefont {David}, \citenamefont {Sinayskiy},\ and\ \citenamefont
  {Petruccione}}]{Pillay2026}%
  \BibitemOpen
  \bibfield  {author} {\bibinfo {author} {\bibfnamefont {S.~M.}\ \bibnamefont
  {Pillay}}, \bibinfo {author} {\bibfnamefont {I.~J.}\ \bibnamefont {David}},
  \bibinfo {author} {\bibfnamefont {I.}~\bibnamefont {Sinayskiy}}, \ and\
  \bibinfo {author} {\bibfnamefont {F.}~\bibnamefont {Petruccione}},\
  }\bibfield  {title} {\enquote {\bibinfo {title} {Optimising trotter-suzuki
  simulations of markovian open quantum systems via classical search},}\ }\href
  {http://dx.doi.org/10.1007/s11128-026-05267-1} {\bibfield  {journal}
  {\bibinfo  {journal} {Quantum Inf. Process.}\ }\textbf {\bibinfo {volume}
  {25}} (\bibinfo {year} {2026})}\BibitemShut {NoStop}%
\bibitem [{\citenamefont {Scopa}\ \emph {et~al.}(2019)\citenamefont {Scopa},
  \citenamefont {Landi}, \citenamefont {Hammoumi},\ and\ \citenamefont
  {Karevski}}]{Scopa2019}%
  \BibitemOpen
  \bibfield  {author} {\bibinfo {author} {\bibfnamefont {S.}~\bibnamefont
  {Scopa}}, \bibinfo {author} {\bibfnamefont {G.~T.}\ \bibnamefont {Landi}},
  \bibinfo {author} {\bibfnamefont {A.}~\bibnamefont {Hammoumi}}, \ and\
  \bibinfo {author} {\bibfnamefont {D.}~\bibnamefont {Karevski}},\ }\bibfield
  {title} {\enquote {\bibinfo {title} {Exact solution of time-dependent
  lindblad equations with closed algebras},}\ }\href {\doibase
  10.1103/PhysRevA.99.022105} {\bibfield  {journal} {\bibinfo  {journal} {Phys.
  Rev. A}\ }\textbf {\bibinfo {volume} {99}},\ \bibinfo {pages} {022105}
  (\bibinfo {year} {2019})}\BibitemShut {NoStop}%
\bibitem [{\citenamefont {Dann}\ \emph {et~al.}(2018)\citenamefont {Dann},
  \citenamefont {Levy},\ and\ \citenamefont {Kosloff}}]{Dann2018}%
  \BibitemOpen
  \bibfield  {author} {\bibinfo {author} {\bibfnamefont {R.}~\bibnamefont
  {Dann}}, \bibinfo {author} {\bibfnamefont {A.}~\bibnamefont {Levy}}, \ and\
  \bibinfo {author} {\bibfnamefont {R.}~\bibnamefont {Kosloff}},\ }\bibfield
  {title} {\enquote {\bibinfo {title} {Time-dependent markovian quantum master
  equation},}\ }\href {\doibase 10.1103/PhysRevA.98.052129} {\bibfield
  {journal} {\bibinfo  {journal} {Phys. Rev. A}\ }\textbf {\bibinfo {volume}
  {98}},\ \bibinfo {pages} {052129} (\bibinfo {year} {2018})}\BibitemShut
  {NoStop}%
\bibitem [{\citenamefont {Wei}\ and\ \citenamefont {Norman}(1963)}]{Wei1963}%
  \BibitemOpen
  \bibfield  {author} {\bibinfo {author} {\bibfnamefont {J.}~\bibnamefont
  {Wei}}\ and\ \bibinfo {author} {\bibfnamefont {E.}~\bibnamefont {Norman}},\
  }\bibfield  {title} {\enquote {\bibinfo {title} {Lie algebraic solution of
  linear differential equations},}\ }\href {\doibase 10.1063/1.1703993}
  {\bibfield  {journal} {\bibinfo  {journal} {J. Math. Phys.}\ }\textbf
  {\bibinfo {volume} {4}},\ \bibinfo {pages} {575--581} (\bibinfo {year}
  {1963})}\BibitemShut {NoStop}%
\bibitem [{\citenamefont {Charzyński}\ and\ \citenamefont
  {Kuś}(2013)}]{Charzynsky2013}%
  \BibitemOpen
  \bibfield  {author} {\bibinfo {author} {\bibfnamefont {S.}~\bibnamefont
  {Charzyński}}\ and\ \bibinfo {author} {\bibfnamefont {M.}~\bibnamefont
  {Kuś}},\ }\bibfield  {title} {\enquote {\bibinfo {title} {{Wei–Norman}
  equations for a unitary evolution},}\ }\href {\doibase
  10.1088/1751-8113/46/26/265208} {\bibfield  {journal} {\bibinfo  {journal}
  {J. Phys. A: Math. Theor.}\ }\textbf {\bibinfo {volume} {46}},\ \bibinfo
  {pages} {265208} (\bibinfo {year} {2013})}\BibitemShut {NoStop}%
\bibitem [{\citenamefont {Wilcox}(1967)}]{Wilcox1967}%
  \BibitemOpen
  \bibfield  {author} {\bibinfo {author} {\bibfnamefont {R.~M.}\ \bibnamefont
  {Wilcox}},\ }\bibfield  {title} {\enquote {\bibinfo {title} {Exponential
  operators and parameter differentiation in quantum physics},}\ }\href
  {\doibase 10.1063/1.1705306} {\bibfield  {journal} {\bibinfo  {journal} {J.
  Math. Phys.}\ }\textbf {\bibinfo {volume} {8}},\ \bibinfo {pages} {962--982}
  (\bibinfo {year} {1967})}\BibitemShut {NoStop}%
\bibitem [{\citenamefont {Suzuki}(1985)}]{Suzuki1985}%
  \BibitemOpen
  \bibfield  {author} {\bibinfo {author} {\bibfnamefont {M.}~\bibnamefont
  {Suzuki}},\ }\bibfield  {title} {\enquote {\bibinfo {title} {Decomposition
  formulas of exponential operators and lie exponentials with some applications
  to quantum mechanics and statistical physics},}\ }\href {\doibase
  10.1063/1.526596} {\bibfield  {journal} {\bibinfo  {journal} {J. Math.
  Phys.}\ }\textbf {\bibinfo {volume} {26}},\ \bibinfo {pages} {601--612}
  (\bibinfo {year} {1985})}\BibitemShut {NoStop}%
\bibitem [{\citenamefont {Suzuki}(1993)}]{Suzuki1993}%
  \BibitemOpen
  \bibfield  {author} {\bibinfo {author} {\bibfnamefont {M.}~\bibnamefont
  {Suzuki}},\ }\bibfield  {title} {\enquote {\bibinfo {title} {General
  decomposition theory of ordered exponentials},}\ }\href {\doibase
  10.2183/pjab.69.161} {\bibfield  {journal} {\bibinfo  {journal} {Proc. Jpn.
  Acad. Ser. B Phys. Biol. Sci.}\ }\textbf {\bibinfo {volume} {69}},\ \bibinfo
  {pages} {161--166} (\bibinfo {year} {1993})}\BibitemShut {NoStop}%
\bibitem [{\citenamefont {Fujii}(2008)}]{Fujii2008}%
  \BibitemOpen
  \bibfield  {author} {\bibinfo {author} {\bibfnamefont {K.}~\bibnamefont
  {Fujii}},\ }\bibfield  {title} {\enquote {\bibinfo {title} {Algebraic
  structure of a master equation with generalized lindblad form},}\ }\href
  {\doibase 10.1142/S0219887808003168} {\bibfield  {journal} {\bibinfo
  {journal} {Int. J. Geom. Methods Mod. Phys.}\ }\textbf {\bibinfo {volume}
  {05}},\ \bibinfo {pages} {1033--1040} (\bibinfo {year} {2008})}\BibitemShut
  {NoStop}%
\bibitem [{\citenamefont {Chru{\'s}ci{\'n}ski}\ and\ \citenamefont
  {Kossakowski}(2010)}]{Chruscinski2010}%
  \BibitemOpen
  \bibfield  {author} {\bibinfo {author} {\bibfnamefont {D.}~\bibnamefont
  {Chru{\'s}ci{\'n}ski}}\ and\ \bibinfo {author} {\bibfnamefont
  {A.}~\bibnamefont {Kossakowski}},\ }\bibfield  {title} {\enquote {\bibinfo
  {title} {General form of quantum evolution},}\ }\href
  {https://arxiv.org/abs/1006.2764} {\  (\bibinfo {year} {2010})},\ \Eprint
  {http://arxiv.org/abs/1006.2764} {arXiv:1006.2764 [quant-ph]} \BibitemShut
  {NoStop}%
\bibitem [{\citenamefont {Ringel}\ and\ \citenamefont
  {Gritsev}(2012)}]{Ringel2012}%
  \BibitemOpen
  \bibfield  {author} {\bibinfo {author} {\bibfnamefont {M.}~\bibnamefont
  {Ringel}}\ and\ \bibinfo {author} {\bibfnamefont {V.}~\bibnamefont
  {Gritsev}},\ }\bibfield  {title} {\enquote {\bibinfo {title} {Liouville
  coherent states},}\ }\href {\doibase 10.1209/0295-5075/99/20012} {\bibfield
  {journal} {\bibinfo  {journal} {Europhys. Lett.}\ }\textbf {\bibinfo {volume}
  {99}},\ \bibinfo {pages} {20012} (\bibinfo {year} {2012})}\BibitemShut
  {NoStop}%
\bibitem [{\citenamefont {Ringel}\ and\ \citenamefont
  {Gritsev}(2013)}]{Ringel2013}%
  \BibitemOpen
  \bibfield  {author} {\bibinfo {author} {\bibfnamefont {M.}~\bibnamefont
  {Ringel}}\ and\ \bibinfo {author} {\bibfnamefont {V.}~\bibnamefont
  {Gritsev}},\ }\bibfield  {title} {\enquote {\bibinfo {title} {Dynamical
  symmetry approach to path integrals of quantum spin systems},}\ }\href
  {\doibase 10.1103/PhysRevA.88.062105} {\bibfield  {journal} {\bibinfo
  {journal} {Phys. Rev. A}\ }\textbf {\bibinfo {volume} {88}},\ \bibinfo
  {pages} {062105} (\bibinfo {year} {2013})}\BibitemShut {NoStop}%
\bibitem [{\citenamefont {Korsch}(2019)}]{Korsch2019}%
  \BibitemOpen
  \bibfield  {author} {\bibinfo {author} {\bibfnamefont {H.~J.}\ \bibnamefont
  {Korsch}},\ }\bibfield  {title} {\enquote {\bibinfo {title} {Lindblad
  dynamics of the damped and forced quantum harmonic oscillator},}\ }\href
  {https://arxiv.org/abs/1908.01187} {\  (\bibinfo {year} {2019})},\ \Eprint
  {http://arxiv.org/abs/1908.01187} {arXiv:1908.01187 [quant-ph]} \BibitemShut
  {NoStop}%
\bibitem [{\citenamefont {Teuber}\ and\ \citenamefont
  {Scheel}(2020)}]{Teuber2020}%
  \BibitemOpen
  \bibfield  {author} {\bibinfo {author} {\bibfnamefont {L.}~\bibnamefont
  {Teuber}}\ and\ \bibinfo {author} {\bibfnamefont {S.}~\bibnamefont
  {Scheel}},\ }\bibfield  {title} {\enquote {\bibinfo {title} {Solving the
  quantum master equation of coupled harmonic oscillators with lie-algebra
  methods},}\ }\href {\doibase 10.1103/PhysRevA.101.042124} {\bibfield
  {journal} {\bibinfo  {journal} {Phys. Rev. A}\ }\textbf {\bibinfo {volume}
  {101}},\ \bibinfo {pages} {042124} (\bibinfo {year} {2020})}\BibitemShut
  {NoStop}%
\bibitem [{\citenamefont {Lin}\ \emph {et~al.}(2025)\citenamefont {Lin},
  \citenamefont {Lapierre}, \citenamefont {Moosavi},\ and\ \citenamefont
  {Ryu}}]{Lin2025}%
  \BibitemOpen
  \bibfield  {author} {\bibinfo {author} {\bibfnamefont {Z.}~\bibnamefont
  {Lin}}, \bibinfo {author} {\bibfnamefont {B.}~\bibnamefont {Lapierre}},
  \bibinfo {author} {\bibfnamefont {P.}~\bibnamefont {Moosavi}}, \ and\
  \bibinfo {author} {\bibfnamefont {S.}~\bibnamefont {Ryu}},\ }\bibfield
  {title} {\enquote {\bibinfo {title} {Chiral instabilities in
  driven-dissipative quantum liquids},}\ }\href {\doibase
  10.1103/PhysRevB.111.165131} {\bibfield  {journal} {\bibinfo  {journal}
  {Phys. Rev. B}\ }\textbf {\bibinfo {volume} {111}},\ \bibinfo {pages}
  {165131} (\bibinfo {year} {2025})}\BibitemShut {NoStop}%
\bibitem [{\citenamefont {Scopa}\ \emph {et~al.}(2018)\citenamefont {Scopa},
  \citenamefont {Landi},\ and\ \citenamefont {Karevski}}]{Scopa2018}%
  \BibitemOpen
  \bibfield  {author} {\bibinfo {author} {\bibfnamefont {S.}~\bibnamefont
  {Scopa}}, \bibinfo {author} {\bibfnamefont {G.~T.}\ \bibnamefont {Landi}}, \
  and\ \bibinfo {author} {\bibfnamefont {D.}~\bibnamefont {Karevski}},\
  }\bibfield  {title} {\enquote {\bibinfo {title} {Lindblad-floquet description
  of finite-time quantum heat engines},}\ }\href {\doibase
  10.1103/PhysRevA.97.062121} {\bibfield  {journal} {\bibinfo  {journal} {Phys.
  Rev. A}\ }\textbf {\bibinfo {volume} {97}},\ \bibinfo {pages} {062121}
  (\bibinfo {year} {2018})}\BibitemShut {NoStop}%
\bibitem [{\citenamefont {Dupays}\ and\ \citenamefont
  {Pain}(2023)}]{Dupays2023}%
  \BibitemOpen
  \bibfield  {author} {\bibinfo {author} {\bibfnamefont {L.}~\bibnamefont
  {Dupays}}\ and\ \bibinfo {author} {\bibfnamefont {J.-C.}\ \bibnamefont
  {Pain}},\ }\bibfield  {title} {\enquote {\bibinfo {title} {Closed forms of
  the zassenhaus formula},}\ }\href {\doibase 10.1088/1751-8121/acc8a3}
  {\bibfield  {journal} {\bibinfo  {journal} {J. Phys. A: Math. Theor.}\
  }\textbf {\bibinfo {volume} {56}},\ \bibinfo {pages} {255202} (\bibinfo
  {year} {2023})}\BibitemShut {NoStop}%
\bibitem [{\citenamefont {Breuer}\ and\ \citenamefont
  {Petruccione}(2002)}]{breuer2002}%
  \BibitemOpen
  \bibfield  {author} {\bibinfo {author} {\bibfnamefont {H.P.}\ \bibnamefont
  {Breuer}}\ and\ \bibinfo {author} {\bibfnamefont {F.}~\bibnamefont
  {Petruccione}},\ }\href {https://books.google.lu/books?id=0Yx5VzaMYm8C}
  {\emph {\bibinfo {title} {The Theory of Open Quantum Systems}}}\ (\bibinfo
  {publisher} {Oxford University Press},\ \bibinfo {year} {2002})\BibitemShut
  {NoStop}%
\bibitem [{\citenamefont {Gyamfi}(2020)}]{Gyamfi_2020}%
  \BibitemOpen
  \bibfield  {author} {\bibinfo {author} {\bibfnamefont {J.~A.}\ \bibnamefont
  {Gyamfi}},\ }\bibfield  {title} {\enquote {\bibinfo {title} {Fundamentals of
  quantum mechanics in liouville space},}\ }\href {\doibase
  10.1088/1361-6404/ab9fdd} {\bibfield  {journal} {\bibinfo  {journal} {Eur. J.
  Phys.}\ }\textbf {\bibinfo {volume} {41}},\ \bibinfo {pages} {063002}
  (\bibinfo {year} {2020})}\BibitemShut {NoStop}%
\bibitem [{\citenamefont {Rivas}\ and\ \citenamefont
  {Huelga}(2012)}]{Rivas_2012}%
  \BibitemOpen
  \bibfield  {author} {\bibinfo {author} {\bibfnamefont {A.}~\bibnamefont
  {Rivas}}\ and\ \bibinfo {author} {\bibfnamefont {S.~F.}\ \bibnamefont
  {Huelga}},\ }\href {\doibase 10.1007/978-3-642-23354-8} {\emph {\bibinfo
  {title} {Open Quantum Systems}}}\ (\bibinfo  {publisher} {Springer Berlin
  Heidelberg},\ \bibinfo {year} {2012})\BibitemShut {NoStop}%
\bibitem [{\citenamefont {Chruściński}\ and\ \citenamefont
  {Kossakowski}(2012)}]{Chruscinski2012}%
  \BibitemOpen
  \bibfield  {author} {\bibinfo {author} {\bibfnamefont {D.}~\bibnamefont
  {Chruściński}}\ and\ \bibinfo {author} {\bibfnamefont {A.}~\bibnamefont
  {Kossakowski}},\ }\bibfield  {title} {\enquote {\bibinfo {title}
  {Markovianity criteria for quantum evolution},}\ }\href {\doibase
  10.1088/0953-4075/45/15/154002} {\bibfield  {journal} {\bibinfo  {journal}
  {J. Phys. B: At. Mol. Opt. Phys.}\ }\textbf {\bibinfo {volume} {45}},\
  \bibinfo {pages} {154002} (\bibinfo {year} {2012})}\BibitemShut {NoStop}%
\bibitem [{\citenamefont {Pinney}(1950)}]{Pinney1950}%
  \BibitemOpen
  \bibfield  {author} {\bibinfo {author} {\bibfnamefont {E.}~\bibnamefont
  {Pinney}},\ }\bibfield  {title} {\enquote {\bibinfo {title} {The nonlinear
  differential equation $y''+p(x)y+cy^{-3}=0$},}\ }\href {\doibase
  10.1090/S0002-9939-1950-0037979-4} {\bibfield  {journal} {\bibinfo  {journal}
  {Proc. Amer. Math. Soc.}\ }\textbf {\bibinfo {volume} {1}},\ \bibinfo {pages}
  {681} (\bibinfo {year} {1950})}\BibitemShut {NoStop}%
\bibitem [{\citenamefont {Gottesman}\ \emph {et~al.}(2001)\citenamefont
  {Gottesman}, \citenamefont {Kitaev},\ and\ \citenamefont
  {Preskill}}]{Gottesman2001}%
  \BibitemOpen
  \bibfield  {author} {\bibinfo {author} {\bibfnamefont {D.}~\bibnamefont
  {Gottesman}}, \bibinfo {author} {\bibfnamefont {A.}~\bibnamefont {Kitaev}}, \
  and\ \bibinfo {author} {\bibfnamefont {J.}~\bibnamefont {Preskill}},\
  }\bibfield  {title} {\enquote {\bibinfo {title} {Encoding a qubit in an
  oscillator},}\ }\href {\doibase 10.1103/PhysRevA.64.012310} {\bibfield
  {journal} {\bibinfo  {journal} {Phys. Rev. A}\ }\textbf {\bibinfo {volume}
  {64}},\ \bibinfo {pages} {012310} (\bibinfo {year} {2001})}\BibitemShut
  {NoStop}%
\bibitem [{\citenamefont {Michael}\ \emph {et~al.}(2016)\citenamefont
  {Michael}, \citenamefont {Silveri}, \citenamefont {Brierley}, \citenamefont
  {Albert}, \citenamefont {Salmilehto}, \citenamefont {Jiang},\ and\
  \citenamefont {Girvin}}]{Michael2016}%
  \BibitemOpen
  \bibfield  {author} {\bibinfo {author} {\bibfnamefont {M.~H.}\ \bibnamefont
  {Michael}}, \bibinfo {author} {\bibfnamefont {M.}~\bibnamefont {Silveri}},
  \bibinfo {author} {\bibfnamefont {R.~T.}\ \bibnamefont {Brierley}}, \bibinfo
  {author} {\bibfnamefont {V.~V.}\ \bibnamefont {Albert}}, \bibinfo {author}
  {\bibfnamefont {J.}~\bibnamefont {Salmilehto}}, \bibinfo {author}
  {\bibfnamefont {L.}~\bibnamefont {Jiang}}, \ and\ \bibinfo {author}
  {\bibfnamefont {S.~M.}\ \bibnamefont {Girvin}},\ }\bibfield  {title}
  {\enquote {\bibinfo {title} {New class of quantum error-correcting codes for
  a bosonic mode},}\ }\href {\doibase 10.1103/PhysRevX.6.031006} {\bibfield
  {journal} {\bibinfo  {journal} {Phys. Rev. X}\ }\textbf {\bibinfo {volume}
  {6}},\ \bibinfo {pages} {031006} (\bibinfo {year} {2016})}\BibitemShut
  {NoStop}%
\bibitem [{\citenamefont {Wang}\ \emph {et~al.}(2026)\citenamefont {Wang},
  \citenamefont {Udupa}, \citenamefont {Hillmann}, \citenamefont {Chabaud},
  \citenamefont {Ferraro},\ and\ \citenamefont {Ferrini}}]{Wang2026}%
  \BibitemOpen
  \bibfield  {author} {\bibinfo {author} {\bibfnamefont {R.}~\bibnamefont
  {Wang}}, \bibinfo {author} {\bibfnamefont {A.}~\bibnamefont {Udupa}},
  \bibinfo {author} {\bibfnamefont {T.}~\bibnamefont {Hillmann}}, \bibinfo
  {author} {\bibfnamefont {U.}~\bibnamefont {Chabaud}}, \bibinfo {author}
  {\bibfnamefont {A.}~\bibnamefont {Ferraro}}, \ and\ \bibinfo {author}
  {\bibfnamefont {G.}~\bibnamefont {Ferrini}},\ }\bibfield  {title} {\enquote
  {\bibinfo {title} {Bosonic quantum error-correcting codes with finite stellar
  rank},}\ }\href {https://arxiv.org/abs/2607.06404} {\  (\bibinfo {year}
  {2026})},\ \Eprint {http://arxiv.org/abs/2607.06404} {arXiv:2607.06404
  [quant-ph]} \BibitemShut {NoStop}%
\bibitem [{\citenamefont {Nielsen}\ and\ \citenamefont
  {Chuang}(2010)}]{Nielsen_Chuang2010}%
  \BibitemOpen
  \bibfield  {author} {\bibinfo {author} {\bibfnamefont {Michael~A.}\
  \bibnamefont {Nielsen}}\ and\ \bibinfo {author} {\bibfnamefont {Isaac~L.}\
  \bibnamefont {Chuang}},\ }\href@noop {} {\emph {\bibinfo {title} {Quantum
  Computation and Quantum Information: 10th Anniversary Edition}}}\ (\bibinfo
  {publisher} {Cambridge University Press},\ \bibinfo {year}
  {2010})\BibitemShut {NoStop}%
\bibitem [{\citenamefont {Steane}(1996)}]{Steane1996}%
  \BibitemOpen
  \bibfield  {author} {\bibinfo {author} {\bibfnamefont {A.~M.}\ \bibnamefont
  {Steane}},\ }\bibfield  {title} {\enquote {\bibinfo {title} {Error correcting
  codes in quantum theory},}\ }\href {\doibase 10.1103/PhysRevLett.77.793}
  {\bibfield  {journal} {\bibinfo  {journal} {Phys. Rev. Lett.}\ }\textbf
  {\bibinfo {volume} {77}},\ \bibinfo {pages} {793--797} (\bibinfo {year}
  {1996})}\BibitemShut {NoStop}%
\bibitem [{\citenamefont {Bennett}\ \emph {et~al.}(1996)\citenamefont
  {Bennett}, \citenamefont {DiVincenzo}, \citenamefont {Smolin},\ and\
  \citenamefont {Wootters}}]{Bennett1996}%
  \BibitemOpen
  \bibfield  {author} {\bibinfo {author} {\bibfnamefont {C.~H.}\ \bibnamefont
  {Bennett}}, \bibinfo {author} {\bibfnamefont {D.~P.}\ \bibnamefont
  {DiVincenzo}}, \bibinfo {author} {\bibfnamefont {J.~A.}\ \bibnamefont
  {Smolin}}, \ and\ \bibinfo {author} {\bibfnamefont {W.~K.}\ \bibnamefont
  {Wootters}},\ }\bibfield  {title} {\enquote {\bibinfo {title} {Mixed-state
  entanglement and quantum error correction},}\ }\href {\doibase
  10.1103/PhysRevA.54.3824} {\bibfield  {journal} {\bibinfo  {journal} {Phys.
  Rev. A}\ }\textbf {\bibinfo {volume} {54}},\ \bibinfo {pages} {3824--3851}
  (\bibinfo {year} {1996})}\BibitemShut {NoStop}%
\bibitem [{\citenamefont {Laflamme}\ \emph {et~al.}(1996)\citenamefont
  {Laflamme}, \citenamefont {Miquel}, \citenamefont {Paz},\ and\ \citenamefont
  {Zurek}}]{Laflamme1996}%
  \BibitemOpen
  \bibfield  {author} {\bibinfo {author} {\bibfnamefont {R.}~\bibnamefont
  {Laflamme}}, \bibinfo {author} {\bibfnamefont {C.}~\bibnamefont {Miquel}},
  \bibinfo {author} {\bibfnamefont {J.~P.}\ \bibnamefont {Paz}}, \ and\
  \bibinfo {author} {\bibfnamefont {W.~H.}\ \bibnamefont {Zurek}},\ }\bibfield
  {title} {\enquote {\bibinfo {title} {Perfect quantum error correcting
  code},}\ }\href {\doibase 10.1103/PhysRevLett.77.198} {\bibfield  {journal}
  {\bibinfo  {journal} {Phys. Rev. Lett.}\ }\textbf {\bibinfo {volume} {77}},\
  \bibinfo {pages} {198--201} (\bibinfo {year} {1996})}\BibitemShut {NoStop}%
\bibitem [{\citenamefont {McDonald}\ and\ \citenamefont
  {Clerk}(2023)}]{McDonald2023}%
  \BibitemOpen
  \bibfield  {author} {\bibinfo {author} {\bibfnamefont {A.}~\bibnamefont
  {McDonald}}\ and\ \bibinfo {author} {\bibfnamefont {A.~A.}\ \bibnamefont
  {Clerk}},\ }\bibfield  {title} {\enquote {\bibinfo {title} {Third
  quantization of open quantum systems: Dissipative symmetries and connections
  to phase-space and keldysh field-theory formulations},}\ }\href {\doibase
  10.1103/PhysRevResearch.5.033107} {\bibfield  {journal} {\bibinfo  {journal}
  {Phys. Rev. Res.}\ }\textbf {\bibinfo {volume} {5}},\ \bibinfo {pages}
  {033107} (\bibinfo {year} {2023})}\BibitemShut {NoStop}%
\bibitem [{\citenamefont {Van-Brunt}\ and\ \citenamefont
  {Visser}(2015)}]{Van-Brunt_2015}%
  \BibitemOpen
  \bibfield  {author} {\bibinfo {author} {\bibfnamefont {A.}~\bibnamefont
  {Van-Brunt}}\ and\ \bibinfo {author} {\bibfnamefont {M.}~\bibnamefont
  {Visser}},\ }\bibfield  {title} {\enquote {\bibinfo {title} {Special-case
  closed form of the baker–campbell–hausdorff formula},}\ }\href {\doibase
  10.1088/1751-8113/48/22/225207} {\bibfield  {journal} {\bibinfo  {journal}
  {J. Phys. A}\ }\textbf {\bibinfo {volume} {48}},\ \bibinfo {pages} {225207}
  (\bibinfo {year} {2015})}\BibitemShut {NoStop}%
\bibitem [{\citenamefont {Dupays}(2025)}]{Dupays2025}%
  \BibitemOpen
  \bibfield  {author} {\bibinfo {author} {\bibfnamefont {L.}~\bibnamefont
  {Dupays}},\ }\bibfield  {title} {\enquote {\bibinfo {title} {Third-quantized
  master equations as a classical ornstein-uhlenbeck process},}\ }\href
  {\doibase 10.1103/ntv6-jzvb} {\bibfield  {journal} {\bibinfo  {journal}
  {Phys. Rev. A}\ }\textbf {\bibinfo {volume} {112}},\ \bibinfo {pages}
  {063724} (\bibinfo {year} {2025})}\BibitemShut {NoStop}%
\end{thebibliography}%

\appendix
\onecolumngrid
\tableofcontents

\section{Derivation of the time-ordered factorization}
This section provides the derivation of Eqs.~\eqref{eq:main_result}-\eqref{eq_anti_chrono} from Suzuki’s operator identity in Sec.~\ref{time_ordering}, followed by a numerical verification in Sec.~\ref{numerical_check}. 
\subsection{Time-ordering and closed Lie algebras \label{time_ordering}}
Here we derive Eqs.~\eqref{eq:main_result}-\eqref{eq_anti_chrono} by substituting $A(s)=\alpha(s)A$ and $B(s)=\beta(s)B$ into Suzuki's formula Eq.~\eqref{eq:Suzuki_equation_p}-\eqref{Suzuki_correction}. This choice leads to the commutation of the adjoint operators at different times
\begin{eqnarray}
C_{\pm}(t)=e^{\mp \Gamma_{\beta}(t,t_{0}) ad_{B} } \left[ e^{ \mp \Gamma_{\alpha}(t,t_{0}) ad_{A} } -1\right]\beta(t)B,
\end{eqnarray}
with $\Gamma_{\alpha}(t,t_{0})=\int_{t_{0}}^{t}ds\;\alpha(s)$ and $\Gamma_{\beta}(t,t_{0})=\int_{t_{0}}^{t}ds\;\beta(s)$. This equation can be rewritten as
\begin{eqnarray}
C_{\pm}(t)=e^{\mp \Gamma_{\beta}(t,t_{0}) ad_{B} } \frac{\left[ e^{ \mp \Gamma_{\alpha}(t,t_{0}) ad_{A} } -1\right]}{ad_{A}}\beta(t)[A,B],\label{eq:coefficient}
\end{eqnarray}
where
\begin{align}
\frac{e^{x ad_{A}}-1}{ad_{A}}=\sum_{n=1}^{\infty}\frac{x^{n}}{n!}ad^{n-1}_{A}.
\end{align}
Using these expressions and the further assumptions $[A,[A,B]]=\lambda_A[A,B]$, $
[B,[A,B]]=\lambda_B[A,B]$ stated in the main text, the correction simplifies, for $\lambda_{A}\neq 0$, to
\begin{eqnarray}
C_{\pm}(t)=e^{\mp \Gamma_{\beta}(t,t_{0}) \lambda_{B} } \frac{\left[ e^{ \mp \Gamma_{\alpha}(t,t_{0}) \lambda_{A} } -1\right]}{\lambda_{A}}\beta(t)[A,B].
\end{eqnarray}
The case $\lambda_{A}=0$ follows by taking the limit in the above expression. Hence, in all cases the correction terms commute at different times $[C_{\pm}(t_{1}),C_{\pm}(t_{2})]=0$ for all $t_{1},t_{2}$ and one can remove the time ordering to obtain Eq.~\eqref{eq:main_result}.
\subsection{Numerical verification of the Lie-algebraic factorizations \label{numerical_check}}
In this section, we perform a numerical verification of our chronological and antichronological Lie-algebraic factorizations, Eq.~\eqref{eq:main_result} and Eq.~\eqref{eq_anti_chrono}. Consider two operators
\begin{align}
X&=\begin{pmatrix}
1 & 0 \\
0 & 0 
\end{pmatrix},& Y&=\begin{pmatrix}
0 & 1\\
0 &0 
\end{pmatrix}.
\end{align}
They satisfy the commutation relation $[X,Y]=Y$. We can define a linear combination
\begin{align}
A&=p X+qY,& B&= r X+s Y. \label{coordinate_transformation}
\end{align}
This leads to 
\begin{align}
[A,B]&=(ps-qr)Y,& [A,[A,B]]&=p[A,B],&[B,[A,B]]&=r[A,B].
\end{align}
Consider now an arbitrary driving protocol of the form 
\begin{align}
\alpha(t)&=0.6 +0.20 \sin( t)\\
\beta(t)&=0.40 +0.15\cos(t ),
\end{align}
for the propagator
\begin{align}
\mathcal{L}(t)&=\alpha(t)A+\beta(t)B.
\end{align}
We can numerically solve for the propagator
\begin{align}
\dot{U}_{+}(t)&=\mathcal{L}(t)U_{+}(t),\quad U_{+}(0)=\mathbbm{1},\\
\dot{U}_{-}(t)&=U_{-}(t)\mathcal{L}(t),\quad U_{-}(0)=\mathbbm{1}.
\end{align}
We compute the relative Frobenius error between the numerical propagator and the analytical propagator. The relative Frobenius error measures the discrepancy between the numerical propagator $U_{\rm num}$ and the factorized propagator $U_{\rm fac}$
\begin{align}
\varepsilon_F(t)
&=
\frac{
\left\| U_{\mathrm{num}}(t)-U_{\mathrm{fac}}(t) \right\|_F
}{
\left\| U_{\mathrm{num}}(t) \right\|_F
},
\end{align}
with the Frobenius norm $\left\| M \right\|_F=\sqrt{{\rm Tr}(M^{\dagger}M)}$. The relative Frobenius error between the numerical and factorized propagators Eq.~\eqref{eq:main_result}-\eqref{eq_anti_chrono} remains negligible for both cases $\lambda_{A} \lambda_{B}\neq 0$ and $\lambda_{A}=0$ with $\lambda_{B}\neq 0$, as displayed in Table~\ref{tab:lie_factorization_numerics}.

\begin{table}[!htbp]
\centering
\caption{Maximum relative Frobenius errors between the numerically integrated propagators and the Lie-algebraic factorizations over $t\in[0,8]$. Form $1$ and form $2$ correspond to the two different orderings displayed respectively for Eq.~\eqref{eq:main_result} and Eq.~\eqref{eq_anti_chrono}. For $\lambda_{A}\neq 0$, we choose $p=0.7$, $q=0.4$,$r=-0.3$,$s=1.1$ for the numerical application for Eq.~\eqref{coordinate_transformation}; for $\lambda_{A}=0$ we choose $p=0$, $q=0.8$,$r=-0.45$,$s=0.6$.}
\label{tab:lie_factorization_numerics}
\begin{tabular}{c cc cccc}
\hline\hline
Case & $\lambda_A$ & $\lambda_B$ & $\mathcal{T}_+$, form 1 & $\mathcal{T}_+$, form 2 & $\mathcal{T}_-$, form 1 & $\mathcal{T}_-$, form 2 \\
\hline
General case & 0.70 & -0.30 & 8.53e-12 & 8.53e-12 & 8.90e-12 & 8.90e-12 \\
$\lambda_A= 0$ & 0.00 & -0.45 & 2.81e-11 & 2.81e-11 & 1.16e-11 & 1.16e-11 \\
\hline\hline
\end{tabular}
\end{table}

\section{Gain--loss propagator and $su(1,1)$ disentangling}
This section derives the exact propagator for the gain--loss master equation. Section~\ref{closed_form_dynamical_map} derives the disentangled propagator and establishes the $su(1,1)$ decomposition. Section~\ref{Wei_Norman_SU11} checks the result by deriving the corresponding Wei--Norman decomposition and verifying that the coefficients $g_{+}(t),g_{0}(t),g_{-}(t)$ satisfy the Wei--Norman system. 

\subsection{Normal-ordered $su(1,1)$ form of the propagator\label{closed_form_dynamical_map}}
In this section, we provide more details on the disentanglement of the dynamical map. The result \eqref{eq:main_result} is applied to the dissipators of annihilation \eqref{dissipator_annihilation} and creation \eqref{dissipator_creation}, with the choice $A=\mathbbm{D}_{a^{\dagger}}$ and $B=\mathbbm{D}_{a}$, such that $\lambda_{A}=1$ and $\lambda_{B}=-1$
\begin{align}
V(t,t_{0})&=\mathcal{T}_{+}\exp\left(\int_{t_{0}}^{t}ds \{\gamma_{+}(s)\mathbbm{D}_{a^{\dagger}}+\gamma_{-}(s)\mathbbm{D}_{a}\}\right)\nonumber\\
&=e^{\Gamma_{+}(t,t_{0})\mathbbm{D}_{a^{\dagger}}}e^{\Gamma_{-}(t,t_{0})\mathbbm{D}_{a}}e^{\Delta_{+}(t,t_{0})[\mathbbm{D}_{a^{\dagger}},\mathbbm{D}_{a}]}\\
&=e^{\Gamma_{+}(t,t_{0})\mathbbm{D}_{a^{\dagger}}}e^{\Delta_{+}(t,t_{0})e^{-\Gamma_{-}(t,t_{0})}[\mathbbm{D}_{a^{\dagger}},\mathbbm{D}_{a}]}e^{\Gamma_{-}(t,t_{0})\mathbbm{D}_{a}},
\end{align}
with 
\begin{align}
\Delta_{+}(t,t_{0};1,-1)&=\int_{t_{0}}^{t}ds\;\left(e^{-\Gamma_{+}(s,t_{0})}-1\right)e^{ \Gamma_{-}(s,t_{0})}\gamma_{-}(s),
\end{align}
with the definition $\Gamma_{\alpha}(t,t_{0})=\int_{t_{0}}^{t}ds\;\gamma_{\alpha}(s)$. We then use the BCH formula for the operators $A,B$ satisfying $[A,B]=uA+vB$ \cite{Van-Brunt_2015}. Hence, for $[\mathbbm{D}_{a},\mathbbm{D}_{a^{\dagger}}]=-(\mathbbm{D}_{a}+\mathbbm{D}_{a^{\dagger}})$, we identify $A=u\mathbbm{D}_{a^{\dagger}}$ and $B=u\mathbbm{D}_{a}$, such that 
\begin{eqnarray}
e^{u\mathbbm{D}_{a^{\dagger}}}e^{u\mathbbm{D}_{a}}=e^{(e^{u}-1)(\mathbbm{D}_{a}+\mathbbm{D}_{a^{\dagger}})}.\label{above_formula}
\end{eqnarray}
By comparing Eq.~\eqref{above_formula} with the central term of Eq.~\eqref{intermed_equation} $e^{u}-1=\Delta_{+}(t,t_{0};1,-1)e^{-\Gamma_{-}(t,t_{0})}$ and $u=\ln\left[\Delta_{+}(t,t_{0};1,-1)e^{-\Gamma_{-}}+1\right]$. For finite Markovian rates ($\gamma_{-}\geq 0$ and $\gamma_{+}\geq 0$), we have 
\begin{align}
\Delta_{+}(t,t_{0};1,-1)&>-\int_{t_{0}}^{t}ds\;e^{\Gamma_{-}(s,t_{0})}\gamma_{-}(s)=-\left(e^{\Gamma_{-}(t,t_{0})}-1\right),\\
1+\Delta_{+}(t,t_{0};1,-1)e^{-\Gamma_{-}(t,t_{0})}&>1-(e^{\Gamma_{-}(t,t_{0})}-1)e^{-\Gamma_{-}(t,t_{0})}=e^{-\Gamma_{-}(t,t_{0})}>0.
\end{align}
Hence, the logarithm is well-defined at finite time $t$. The limit $\Gamma_{-}(t,t_{0})\to \infty$ requires careful treatment. However, the limit depends on the asymptotic behavior of the rates, and can only be treated for each particular case. We can pursue the decomposition as
\begin{align}
V(t,t_{0})&=e^{\Phi_{+}\mathbbm{D}_{a^{\dagger}}}e^{\Phi_{-}\mathbbm{D}_{a}},\\
\Phi_{+}&=\Gamma_{+}(t,t_{0})+\ln[\Delta_{+}(t,t_{0};1,-1)e^{-\Gamma_{-}}+1],\\
\Phi_{-}&=\Gamma_{-}(t,t_{0})+\ln[\Delta_{+}(t,t_{0};1,-1)e^{-\Gamma_{-}}+1].
\end{align}
Further decomposition in terms of the operators of the $su(1,1)$ algebra $K_{+}$, $K_{-}$ and $K_{0}$ is now detailed. Let us recall the formula \cite{Dupays2023} for two operators satisfying $[A,B]=uA+vB$
\begin{eqnarray}
\exp(A+B)&=&\exp(A)\exp(B)\exp(g_{r}(u,v)[A,B]),\label{Z_formula}\\
&=&\exp(g_{l}(u,v)[A,B])\exp(A)\exp(B),\\
g_{r}(0,v)&=&-\frac{e^{-v}-1+v}{v^{2}},\\
g_{r}(v,0)&=&\frac{e^{v}(1-v)-1}{v^{2}},\\
g_{l}(u,v)&=&g_{r}(v,u).
\end{eqnarray}
Following the decomposition of the dissipator operators in terms of the $su(1,1)$ algebra components given in the main text, that we recall here for convenience $\mathbbm{D}_{a}=K_{-}-K_{0}+\frac{1}{2}\mathbbm{1}\otimes \mathbbm{1}, \mathbbm{D}_{a^{\dagger}}=K_{+}-K_{0}-\frac{1}{2}\mathbbm{1}\otimes \mathbbm{1}$,  one can apply the Zassenhaus formula \eqref{Z_formula} and obtain the decompositions
\begin{align}
e^{\Phi \mathbbm{D}_{a}}&=e^{-\Phi K_{0}}e^{\Phi[1+\Phi g_{r}(0,\Phi)]K_{-}}e^{\frac{\Phi}{2}(\mathbbm{1}\otimes \mathbbm{1})},\\
e^{\Phi \mathbbm{D}_{a^{\dagger}}}&=e^{\Phi [1+\Phi g_{l}(\Phi,0)]K_{+}}e^{-\Phi K_{0}}e^{-\frac{\Phi }{2}(\mathbbm{1}\otimes \mathbbm{1})}.
\end{align}
Merging the two decompositions, one obtains the final result
\begin{align}
e^{\Phi_{+}\mathbbm{D}_{a^{\dagger}}}e^{\Phi_{-}\mathbbm{D}_{a}}&=e^{g_{+}K_{+}}e^{g_{0}K_{0}}e^{g_{-}K_{-}}e^{\frac{\Phi_{-}-\Phi_{+}}{2}},\\
g_{+}&=\Phi_{+}[1+\Phi_{+}g_{l}(\Phi_{+},0)]=1-e^{-\Phi_{+}},\\
g_{-}&=\Phi_{-}[1+\Phi_{-}g_{r}(0,\Phi_{-})]=1-e^{-\Phi_{-}},\\
g_{0}&=-(\Phi_{+}+\Phi_{-}),
\end{align}
which is also given in Eq.~\eqref{SU11_decomposition} of the main text. The final result is presented in the main text Eq.~\eqref{SU11_decomposition}-\eqref{delta_plus}.
\subsection{Wei--Norman decomposition consistency check \label{Wei_Norman_SU11}}
We would like to verify that the $su(1,1)$ decomposition Eq.~\eqref{SU11_decomposition} is a solvable Wei--Norman decomposition, in the sense that we can determine an explicit expression for the coefficients $g_{-},g_{+},g_{0}$. Let us establish the Wei--Norman decomposition for the dynamical map. Consider the logarithmic derivative
\begin{align}
\dot{V}V^{-1}&=\dot{g}_{+}K_{+}+\dot{g}_{0}e^{g_{+}{\rm ad}_{K_{+}}}K_{0}+\dot{g}_{-}e^{g_{+}{\rm ad}_{K_{+}}}e^{g_{0}{\rm ad}_{K_{0}}}K_{-}+\frac{\dot{\Phi}_{-}-\dot{\Phi}_{+}}{2}\mathbbm{1}.
\end{align}
Let us recall the $su(1,1)$ algebra
\begin{align}
[K_{0},K_{\pm}]&=\pm K_{\pm},\quad [K_{-},K_{+}]=2K_{0}.
\end{align}
Hence, we obtain
\begin{align}
e^{g_{+}{\rm ad}_{K_{+}}}K_{0}&=K_{0}-g_{+}K_{+},\\
e^{g_{0}{\rm ad}_{K_{0}}}K_{-}&=e^{-g_{0}}K_{-},\\
e^{g_{+}{\rm ad}_{K_{+}}}K_{-}&=K_{-}-2g_{+}K_{0}+g^{2}_{+}K_{+}.
\end{align}
Hence, 
\begin{align}
\dot{V}V^{-1}&=K_{0}(\dot{g}_{0}-2g_{+}\dot{g}_{-}e^{-g_{0}})+K_{+}(\dot{g}_{+}-\dot{g}_{0}g_{+}+\dot{g}_{-}g^{2}_{+}e^{-g_{0}})+K_{-}e^{-g_{0}}\dot{g}_{-}+\frac{\dot{\Phi}_{-}-\dot{\Phi}_{+}}{2}\mathbbm{1}.\label{VV_minus}
\end{align}
Furthermore, we have
\begin{align}
    \gamma_{+}(t)\mathbbm{D}_{a^{\dagger}}+\gamma_{-}(t)\mathbbm{D}_{a}
    &=-K_{0}[\gamma_{+}(t)+\gamma_{-}(t)]+\gamma_{+}(t)K_{+}+\gamma_{-}(t)K_{-}+\frac{1}{2}[\gamma_{-}(t)-\gamma_{+}(t)].\label{gain_loss_dissipator}
\end{align}
Matching the coefficients of $K_{0},K_{+},K_{-}$ for Eq.~\eqref{VV_minus} with Eq.~\eqref{gain_loss_dissipator} gives the system of differential equations Eq.~\eqref{WeiNorman_g+}-~\eqref{WeiNorman_g-}. The above system of differential equations is solvable. Our disentangling procedure yields the solution. One can verify that the solution presented in Eq.~\eqref{Phi_-}-\eqref{g_-} solves the previous system of differential equations. Introducing
\begin{align}
Q&=1+\Delta_{+}(t,t_{0};1,-1)e^{-\Gamma_{-}}.
\end{align}
One finds
\begin{align}
\dot{Q}&=\gamma_{-}(e^{-\Gamma_{+}}-Q).
\end{align}
Further substituting, for real $Q\neq 0$
\begin{align}
g_{+}&=1-\frac{e^{-\Gamma_{+}}}{Q},& g_{-}&=1-\frac{e^{-\Gamma_{-}}}{Q},& e^{g_{0}}&=\frac{e^{-(\Gamma_{+}+\Gamma_{-})}}{Q^{2}},
\end{align}
into the Wei--Norman equations verifies them identically 
\begin{align}
\dot{g}_{+}&=\gamma_{+}\frac{e^{-\Gamma_{+}}}{Q}+\frac{\dot{Q}}{Q^{2}}e^{-\Gamma_{+}}\nonumber=\gamma_{+}-(\gamma_{-}+\gamma_{+})g_{+}+\gamma_{-}g^{2}_{+}.
\end{align}
And
\begin{align}
\dot{g}_{-}&=\gamma_{-}\frac{e^{-\Gamma_{-}}}{Q}+\frac{\dot{Q}}{Q^{2}}e^{-\Gamma_{-}}=\gamma_{-}e^{g_{0}}.
\end{align}
Finally, 
\begin{align}
\dot{g}_{0}&=-(\gamma_{+}+\gamma_{-})-2\frac{\dot{Q}}{Q}=-(\gamma_{+}+\gamma_{-})+2\gamma_{-}g_{+}.
\end{align}
At $t=t_{0}$, we have $\Gamma_{+}(t_{0})=\Gamma_{-}(t_{0})=\Delta_{+}(t_{0},t_{0};1,-1)=0$. Hence, $g_{+}(t_{0})=g_{-}(t_{0})=g_{0}(t_{0})=0$.
\section{Fock-state dynamics and its Wigner function}
This section focuses on the dynamics of an initial Fock state and its associated Wigner function. Section~\ref{propagator_Fock} derives the dynamics of the density-matrix elements in the Fock basis. Then, Section~\ref{verification_birth_death} verifies the dynamics of the diagonal Fock state by comparison with an independent solution of the birth--death master equation. Section~\ref{Wigner_dynamics} derives the Wigner function dynamics for the diagonal Fock state. Section~\ref{verification_Wigner_function} proposes an independent phase-space verification of the Wigner function dynamics. 

\subsection{Action of the propagator on the Fock basis \label{propagator_Fock}}
The operators $K_{0}$ \eqref{eq:K0}, $K_{+}$ \eqref{eq:K+} and $K_{-}$ \eqref{eq:K-} given in the main text obey the $su(1,1)$ canonical commutation relations $[K_{0},K_{\pm}]=\pm K_{\pm}$ and $[K_{-},K_{+}]=2K_{0}$. These operators are ladder operators for the Fock basis 
\begin{align}
K_{0}|n,m\rrangle&=\frac{1}{2}(n+m+1)|n,m\rrangle,\\
K_{+}|n,m\rrangle&=\sqrt{(n+1)(m+1)}|n+1,m+1\rrangle,\\
K_{-}|n,m\rrangle&=\sqrt{n m}|n-1,m-1\rrangle.\\
\end{align}
Here, $|n,m\rrangle=|n\rangle \otimes |m\rangle$. The action of the generators of the algebra reads
\begin{align}
e^{g_{+}K_{+}}|n,m\rrangle&=\sum_{p=0}^{\infty}(g_{+})^{p}\sqrt{\binom{n+p}{n}\binom{m+p}{m}}|n+p,m+p\rrangle,\\
e^{g_{0}K_{0}}|n,m\rrangle&=e^{\frac{1}{2}(n+m+1)g_{0}}|n,m\rrangle,\\
e^{g_{-}K_{-}}|n,m\rrangle&=\sum_{p=0}^{{\rm min}(n,m)}(g_{-})^{p}\sqrt{\binom{n}{p}\binom{m}{p}}|n-p,m-p\rrangle.
\end{align}
Finally, the generalized Fock state evolves as
\begin{align}
e^{g_{+}K_{+}}e^{g_{0}K_{0}}e^{g_{-}K_{-}}|n,m\rrangle=\sum_{j=0}^{\infty}\sum_{p=0}^{{\rm min}(n,m)}e^{\frac{1}{2}(n+m-2p+1)g_{0}}(g_{-})^{p}(g_{+})^{j}\sqrt{\binom{n}{p}\binom{m}{p}\binom{n-p+j}{n-p}\binom{m-p+j}{m-p}}|n-p+j,m-p+j\rrangle.
\end{align}
\subsection{Independent birth--death verification \label{verification_birth_death}}
This section provides an independent verification of the Fock-state propagator. For an initial Fock state $|n,n\rrangle$, the disentangled $su(1,1)$ propagator reduces to 
\begin{align}
e^{g_{+}K_{+}}e^{g_{0}K_{0}}e^{g_{-}K_{-}}|n,n\rrangle=\sum_{j=0}^{\infty}\sum_{p=0}^{n}e^{(n-p+\frac{1}{2})g_{0}}(g_{-})^{p}(g_{+})^{j}\binom{n}{p}\binom{n-p+j}{n-p}|n-p+j,n-p+j\rrangle.\label{Fock_state_evolution}
\end{align}
We now verify the Fock-state decomposition in Eq.~\eqref{Fock_state_evolution}. The verification is performed by comparing two independently derived generating functions. The first is obtained directly from the probability distribution generated by the $su(1,1)$ propagator, while the second follows from the birth--death equation associated with the gain--loss master equation. Consider the generating function
\begin{align}
G(z,t)&=\sum_{n}p_{n}(t)z^{n},\label{generating_function}
\end{align}
associated with $p_{n}(t)$, the probability of occupying the $n$-th Fock state and $z$ an auxiliary variable. Equality of two generating functions for all values of the auxiliary variable $z$ implies equality of all Fock-state populations.
Writing the final Fock-state index as $r=n-p+j$, the corresponding population is therefore 
\begin{align}
p_{r}(t)&=\mathcal{N}(t)\sum_{p={\rm max}(0,n-r)}^{n}e^{(n-p+\frac{1}{2})g_{0}}(g_{-})^{p}(g_{+})^{r-n+p}\binom{n}{p}\binom{r}{n-p},
\end{align}
where $\mathcal{N}(t)=e^{\frac{\Phi_{-}-\Phi_{+}}{2}}$ denotes the scalar prefactor multiplying the $su(1,1)$ propagator. The generating function associated with this probability distribution is defined as
\begin{align}
G_{SU}(z,t)&=\sum_{r=0}^{\infty}p_{r}(t)z^{r}.
\end{align}
One can construct the generating function directly from Eq.~\eqref{Fock_state_evolution}
\begin{align}
G_{SU}(z,t)&=\mathcal{N}(t)\sum_{p=0}^{n}e^{(n-p+\frac{1}{2})g_{0}}(g_{-})^{p}\binom{n}{p}z^{n-p}\sum_{j=0}^{\infty}\binom{n-p+j}{n-p}(g_{+}z)^{j}.
\end{align}
Using the identity 
\begin{align}
\sum_{j=0}^{\infty}\binom{k+j}{k}y^{j}&=\frac{1}{(1-y)^{k+1}},\quad |y|<1,
\end{align}
with $k=n-p$, and $y=g_{+}z$, we find
\begin{align}
G_{SU}(z,t)&=\mathcal{N}(t)\frac{e^{g_{0}/2}}{1-g_{+}z}\sum_{p=0}^{n}\binom{n}{p}g^{p}_{-}\left[\frac{e^{g_{0}}z}{1-g_{+}z}\right]^{n-p}.
\end{align}
The binomial theorem then yields
\begin{align}
G_{SU}(z,t)
&=\mathcal{N}(t)e^{g_{0}/2}\frac{[g_{-}+(e^{g_{0}}-g_{-}g_{+})z]^{n}}{(1-g_{+}z)^{n+1}}.\label{generating_function_SU11}
\end{align}
We next derive the same generating function independently. Consider an initial state diagonal in the Fock basis
\begin{align}
\rho(0)&=\sum_{n}p_{n}(0)|n\rangle\langle n|.
\end{align}
Its evolution is governed by the gain--loss master equation Eq.~\eqref{Lindblad_master_equation}, which preserves diagonality in the Fock basis. We may therefore write 
\begin{align}
\rho(t)&=\sum_{n}p_{n}(t)|n\rangle\langle n|.
\end{align}
Consequently, the populations satisfy the birth--death equation
\begin{align}
\dot{p}_{n}&=\gamma_{-}(t)[(n+1)\;p_{n+1}-n\;p_{n}]+\gamma_{+}(t)[n\;p_{n-1}-(n+1)\;p_{n}],\label{eq:birth_death}
\end{align}
with the boundary convention $p_{-1}(t)=0$. To solve Eq.~\eqref{eq:birth_death}, we can solve for the generating function Eq.~\eqref{generating_function}. Multiplying Eq.~\eqref{eq:birth_death} by $z^{n}$ and summing over $n\geq 0$, we obtain
\begin{align}
\sum_{n=0}^{\infty}\dot{p}_{n}z^{n}&=\gamma_{-}(t)\Big[\sum_{n=0}^{\infty}(n+1)p_{n+1}z^{n}-\sum_{n=0}^{\infty}n p_{n}z^{n}\Big]+\gamma_{+}(t)\Big[\sum_{n=0}^{\infty}n p_{n-1}z^{n}-\sum_{n=0}^{\infty}(n+1) p_{n}z^{n}\Big].
\end{align}
The sums can be written
\begin{align}
\sum_{n=0}^{\infty}(n+1)p_{n+1}z^{n}&=\partial_{z}G,& \sum_{n=0}^{\infty}n p_{n}z^{n}&=z\partial_{z}G,\\
\sum_{n=0}^{\infty}np_{n-1}z^{n}&=z^{2}\partial_{z}G+zG,& \sum_{n=0}^{\infty}(n+1)p_{n}z^{n}&=z\partial_{z}G+G.
\end{align}
It follows that 
\begin{align}
    \partial_{t}G(z,t)&=(1-z)[\gamma_{-}(t)-z\gamma_{+}(t)]\partial_{z}G(z,t)-\gamma_{+}(t)(1-z)G(z,t). \label{equation_characteristics}
\end{align}
We solve Eq.~\eqref{equation_characteristics} by the method of characteristics. This corresponds to choosing a trajectory $u(t)$ such that the total derivative of $G$ along that trajectory reproduces the two derivative terms
\begin{align}
\frac{dG}{dt}&=\partial_{t}G+\dot{u}(t)\partial_{u}G.
\end{align}
Introduce
\begin{align}
u&=1-z,\\
\delta(t)&=\gamma_{-}(t)-\gamma_{+}(t).\label{delta_definition}
\end{align}
Since $\partial_{z}=-\partial_{u}$, Eq.~\eqref{equation_characteristics} becomes
\begin{align}
\partial_{t}G(u,t)+u[\delta(t)+u\gamma_{+}(t)]\partial_{u}G(u,t)+\gamma_{+}u G(u,t)&=0.
\end{align}
The corresponding characteristic equations are by identification
\begin{align}
\frac{du}{dt}&=u[\delta(t)+\gamma_{+}(t)u], \label{riccati_characteristic}\\
\frac{d G}{dt}&=-\gamma_{+}(t)u(t)G.\label{characteristic_G}
\end{align}
Eq.~\eqref{riccati_characteristic} is a Riccati equation. To solve it, let us introduce
\begin{align}
w(t)&=\frac{1}{u(t)}.
\end{align}
We obtain the linear equation
\begin{align}
\dot{w}(t)+\delta(t)w(t)&=-\gamma_{+}(t).
\end{align}
Define
\begin{align}
K(t)&=\int_{0}^{t}ds\;\delta(s)=\Gamma_{-}(t,0)-\Gamma_{+}(t,0).\label{K_definition}
\end{align}
Multiplying by the integrating factor $e^{K(t)}$ gives
\begin{align}
\frac{d}{dt}[e^{K(t)}w(t)]&=-e^{K(t)}\gamma_{+}(t).
\end{align}
Integrating from $0$ to $t$, we find
\begin{align}
e^{K(t)}w(t)-w_{0}&=-\int_{0}^{t}ds\;e^{K(s)}\gamma_{+}(s),
\end{align}
where $w_{0}=w(0)$. Therefore
\begin{align}
\frac{1}{u(t)}&=\frac{e^{-K(t)}}{u_{0}}-e^{-K(t)}\int_{0}^{t}ds\;e^{K(s)}\gamma_{+}(s).
\end{align}
We further introduce
\begin{align}
\eta(t)&=e^{-K(t)},\label{eta_definition}\\
\nu(t)&=e^{-K(t)}\int_{0}^{t}ds\; e^{K(s)}\gamma_{+}(s)=\int_{0}^{t}ds\; e^{-[K(t)-K(s)]}\gamma_{+}(s).\label{nu_definition}
\end{align}
The characteristic equation can then be written as
\begin{align}
\frac{1}{u}&=\frac{\eta(t)}{u_{0}}-\nu(t),
\end{align}
or equivalently, solving for $u_{0}$ gives
\begin{align}
u_{0}&=\frac{\eta(t)u}{1+\nu(t)u}.\label{eq_u0}
\end{align}
Since $z_{0}=1-u_{0}$, Eq.~\eqref{eq_u0} gives
\begin{align}
z_{0}&=1-\frac{\eta(t)(1-z)}{1+\nu(t)(1-z)}.
\end{align}
Combining Eq.~\eqref{riccati_characteristic} and Eq.~\eqref{characteristic_G}
\begin{align}
\frac{d}{dt}\ln[G u ]&=\delta(t).
\end{align}
After integration,
\begin{align}
G(z,t)u(t)&=e^{K(t)}G(z_{0},0)u_{0}.
\end{align}
Thus,
\begin{align}
G(z,t)&=\frac{1}{1+\nu(t)u}G\left(1-\frac{\eta(t)(1-z)}{1+\nu(t)(1-z)},0\right).
\end{align}
For an initial Fock state $|n\rangle\langle n|$, the generating function is
\begin{align}
G(z,0)&=z^{n}.
\end{align}
Hence, the birth--death generating function is 
\begin{align}
G_{\rm BD}(z,t)&=\frac{[1+(\nu-\eta)(1-z)]^{n}}{[1+\nu(1-z)]^{n+1}}.\label{generating_birth_death}
\end{align}
The two expressions Eq.~\eqref{generating_birth_death} and Eq.~\eqref{generating_function_SU11} coincide upon identifying
\begin{align}
    g_{+}&=\frac{\nu}{1+\nu},& g_{-}&=\frac{1+\nu-\eta}{1+\nu},& e^{g_{0}}&=\frac{\eta}{(1+\nu)^{2}}.\label{g0_nu}
\end{align}
Indeed, these relations imply 
\begin{align}
1-g_{+}z&=\frac{1+\nu(1-z)}{1+\nu},\\
g_{-}+(e^{g_{0}}-g_{-}g_{+})z&=\frac{1+(\nu-\eta)(1-z)}{1+\nu}.
\end{align}
Consequently, equality of the two generating functions requires
\begin{align}
\mathcal{N}(t)&=\frac{e^{-g_{0}/2}}{1+\nu}=\eta^{-1/2}=e^{\frac{1}{2}[\Gamma_{-}(t,0)-\Gamma_{+}(t,0)]}.
\end{align}
This is precisely the scalar normalization factor to recover the dynamical map $V(t,0)$. Therefore, 
\begin{align}
G_{SU}(z,t)&=G_{\rm BD}(z,t).
\end{align}
Since the generating functions coincide for all $z$, their Taylor coefficients are identical. It follows that the $su(1,1)$ propagator reproduces the complete Fock-state population dynamics generated by the birth--death equation. By linearity, the result extends to any initial state diagonal in the Fock basis. We have demonstrated that the Fock state evolution we obtained from the disentangled propagator coincides with the evolution obtained from the birth--death master equation. 
\subsection{Wigner function dynamics \label{Wigner_dynamics}}
The initial Wigner function for a pure state is given by
\begin{align}
W_{|n\rangle}(x,p)=\frac{(-1)^{n}}{\pi}e^{-R}L_{n}(2R),
\end{align}
with $R\equiv x^{2}+p^{2}$. Also, as the Fock state evolution can be described in the Fock state basis from \eqref{Fock_state_evolution}, one can write the evolution of the Wigner function as
\begin{align}
W_{|n\rangle}(x,p,t)&=\mathcal{N}(t)\sum_{j=0}^{\infty}\sum_{k=0}^{n}e^{(n-k+\frac{1}{2})g_{0}}(g_{-})^{k}(g_{+})^{j}\binom{n}{k}\binom{n-k+j}{n-k}W_{|n-k+j\rangle}(x,p,0)\\
&=\mathcal{N}(t)\frac{e^{-R}}{\pi}\sum_{k=0}^{n}(-1)^{n-k}e^{(n-k+\frac{1}{2})g_{0}}(g_{-})^{k}\binom{n}{k}\sum_{j=0}^{\infty}(-g_{+})^{j}\binom{n-k+j}{n-k}L_{n-k+j}(2R).\label{intermed_wigner}
\end{align}
It is now required to evaluate 
\begin{align}
J_{k}(g,y)&=\sum_{j=0}^{\infty}g^{j}\binom{k+j}{k}L_{k+j}(y).
\end{align}
We use the integral representation of the Laguerre polynomial through the Cauchy formula
\begin{align}
L_{n}(y)&=\frac{1}{2\pi i}\oint dt \frac{e^{-yt/(1-t)}}{(1-t)t^{n+1}}.
\end{align}
Hence, we obtain
\begin{align}
J_{k}(g,y)&=\sum_{j=0}^{\infty}\binom{k+j}{k}g^{j}\frac{1}{2\pi i}\oint dt \frac{e^{-yt/(1-t)}}{(1-t)t^{k+j+1}}=\frac{1}{2\pi i}\oint dt\; \frac{e^{-yt/(1-t)}}{(1-t)t^{k+1}}\sum_{j=0}^{\infty}\binom{k+j}{k}\left(\frac{g}{t}\right)^{j}.
\end{align}
Now use
\begin{align}
\sum_{j=0}^{\infty}\binom{k+j}{k}y^{j}=\frac{1}{(1-y)^{k+1}}, \quad |y|<1.
\end{align}
Then
\begin{align}
J_{k}(g,y)&=\frac{1}{2\pi i}\oint dt\; \frac{e^{-yt/(1-t)}}{(1-t)\left(t-g\right)^{k+1}}.
\end{align}
Let us make the affine change of variable
\begin{align}
t=g+(1-g)u,
\end{align}
so that $dt=(1-g)du$. One obtains
\begin{align}
\frac{t}{1-t}&=\frac{g}{1-g}+\frac{u}{(1-g)(1-u)}.
\end{align}
By substitution
\begin{align}
J_{k}(g,y)&=\frac{e^{-yg/(1-g)}}{(1-g)\left(1-g\right)^{k+1}}\frac{1}{2\pi i}\oint (1-g)du\; \frac{e^{-\frac{yu}{(1-g)(1-u)}}}{(1-u)u^{k+1}}\\
&=\frac{1}{\left(1-g\right)^{k+1}}\exp\left(-\frac{yg}{1-g}\right)L_{k}\left(\frac{y}{1-g}\right).
\end{align}
Hence, the Wigner function of Eq.~\eqref{intermed_wigner} is further expressed as
\begin{align}
W_{|n\rangle}(x,p,t)&=\mathcal{N}(t)\frac{1}{\pi}\exp\left(\frac{2R g_{+}}{1+g_{+}}\right)e^{-R}\frac{(-e^{g_{0}})^{n}e^{\frac{g_{0}}{2}}}{(1+g_{+})^{n+1}}\sum_{k=0}^{n}\binom{n}{k}[-e^{-g_{0}}g_{-}(1+g_{+})]^{k}L_{n-k}\left(\frac{2R}{1+g_{+}}\right).
\end{align}
To further simplify this expression, one needs to determine the generating function associated with the polynomial in the infinite sum. We introduce the polynomials of order $k$
\begin{align}
D_{k}(g,y)=\sum_{j=0}^{k}\binom{k}{j}g^{j}L_{k-j}(y). \label{D_polynomial}
\end{align}
To determine its generating function, one can introduce the integral representation of the Laguerre polynomials through the Cauchy formula
\begin{align}
L_{n}(y)&=\frac{1}{2\pi i}\oint dt \frac{e^{-yt/(1-t)}}{(1-t)t^{n+1}}.
\end{align}
Inserting the previous expression in the polynomial, one obtains
\begin{align}
D_{k}(g,y)&=\frac{1}{2\pi i}\oint dt \frac{e^{-yt/(1-t)}}{(1-t)t^{k+1}}\sum_{j=0}^{k}\binom{k}{j}g^{j}t^{j}\\
&=\frac{1}{2\pi i}\oint dt \frac{e^{-yt/(1-t)}(1+gt)^{k}}{(1-t)t^{k+1}}.
\end{align}
Assuming that $g\neq -1$, it is now convenient to make the change of variable $t'=t/(1+gt)$ and $y'=y/(g+1)$, so that one can recast the polynomial as
\begin{align}
D_{k}(g,y)&=(g+1)^{k+1}\frac{1}{2\pi i}\oint dt' \frac{e^{-\frac{y'(g+1)t'}{1-(g+1)t'}}}{[1-(g+1)t'][(g+1)t']^{k+1}},\\
&=(g+1)^{k}L_{k}\left(\frac{y}{g+1}\right),
\end{align}
where the last line is obtained by the change of variable $u=(g+1)t'$. Finally, we obtain
\begin{align}
W_{|n\rangle}(x,p,t)&=\mathcal{N}(t)\frac{(-1)^{n}}{\pi}\exp\left(-R\frac{1-g_{+}}{1+g_{+}}\right)\frac{e^{(n+\frac{1}{2})g_{0}}}{(1+g_{+})^{n+1}}\Lambda^{n}L_{n}\left[\frac{2 R }{(1+g_{+})\Lambda}\right],
\end{align}
with $R=x^{2}+p^{2}$ and $\Lambda=1-e^{-g_{0}}g_{-}(1+g_{+})$. And
\begin{align}
\mathcal{N}(t)=(1-g_{+})e^{-g_{0}/2}.
\end{align}
At time $t=t_{0}$, we have $g_{+}(t_{0})=g_{-}(t_{0})=g_{0}(t_{0})=0$, and the Wigner function reduces to  
\begin{align}
W_{|n\rangle}(x,p,0)&=\frac{(-1)^{n}}{\pi}\exp(-R)L_{n}\left(2R\right).
\end{align}
This is the expected result. 

Also consider the pure--loss limit 
$g_{+}=0, g_{-}=1-\upsilon, g_{0}=\ln(\upsilon),$
with $\upsilon=e^{-\Gamma_{-}(t,t_{0})}$. In this situation the Wigner function reduces to 
\begin{align}
W_{|n\rangle}(x,p,t)&=\frac{(-1)^{n}}{\pi}\exp(-R)(2\upsilon-1)^{n} L_{n}\left(2 R\frac{\upsilon}{2\upsilon-1}\right).
\end{align}
There is a singularity at $\upsilon=\frac{1}{2}$, however it can be removed from
\begin{align}
{\rm lim}_{\upsilon\to \frac{1}{2}}(2\upsilon-1)^{n}L_{n}\left(\frac{2\upsilon R}{2\upsilon-1}\right)=\frac{(-R)^{n}}{n!}.
\end{align}
\subsection{Independent phase-space verification \label{verification_Wigner_function}}
This section provides an independent check for the Wigner function evolution. The Wigner function obtained from the normal-ordered $su(1,1)$ propagator can be verified independently by solving the phase-space equation associated with the same gain--loss master equation. Consider the gain--loss master equation Eq.~\eqref{Lindblad_master_equation}. Using the Wigner-Weyl correspondence, the phase-space representation of the master equation is \cite{Dupays2025}
\begin{align}
\partial_{t}W&=\frac{\delta(t)}{2}\Big[\partial_{\alpha}(\alpha W)+\partial_{\alpha^{*}}(\alpha^{*}W)\Big]+[\gamma_{-}(t)+\gamma_{+}(t)]\partial_{\alpha}\partial_{\alpha^{*}}W,
\end{align}
with the $\delta (t)$ of Eq.~\eqref{delta_definition} and the conventions for the Wigner function and its associated characteristic function
\begin{align}
W(\alpha)&=2{\rm Tr}[e^{i\pi a^{\dagger}a}D^{\dagger}(\sqrt{2}\alpha)\rho],\\
\chi_{s}(\lambda,\lambda^{*})&=\int \frac{d^{2}\alpha}{2\pi}e^{\lambda^{*}\alpha-\lambda \alpha^{*}}W(\alpha),
\end{align}
and the displacement operator is defined as $\hat{D}(\alpha)\equiv e^{\alpha a^{\dagger}-\alpha^{*}a}$. The characteristic function equation obeys
\begin{align}
\partial_{t}\chi_{s}&=-\frac{\delta(t)}{2}(\lambda \partial_{\lambda}+\lambda^{*}\partial_{\lambda^{*}})\chi_{s}-[\gamma_{-}(t)+\gamma_{+}(t)]|\lambda|^{2}\chi_{s}.
\end{align}
This equation is a first-order partial differential equation and is integrable. One can solve the previous equation by the method of characteristics and obtain
\begin{align}
\chi_{s}(\lambda,\lambda^{*},t)&=\chi_{s}\left(\sqrt{\eta(t)}\lambda,\sqrt{\eta(t)}\lambda^{*},0\right)\exp\big\{-\big[1-\eta(t)+2\nu(t)\big]|\lambda|^{2}\big\},
\end{align}
where we used the previous definitions for $K(t)$ Eq.~\eqref{K_definition}, $\nu(t)$ Eq.~\eqref{nu_definition}, and $\eta(t)$ Eq.~\eqref{eta_definition}. The initial Fock-state characteristic function is well-known
\begin{align}
\chi_{s}(\lambda,\lambda^{*},0)&={\rm Tr}\Big(D^{\dagger}(\sqrt{2}\lambda)|n\rangle\langle n|\Big)=\langle n|e^{\sqrt{2}\lambda^{*}a-\sqrt{2}\lambda a^{\dagger}}|n\rangle=e^{-|\lambda|^{2}}L_{n}(2|\lambda|^{2}).
\end{align}
Hence, we obtain the characteristic evolution for the Fock state
\begin{align}
\chi_{s}(\lambda,\lambda^{*},t)&=e^{-\eta(t)|\lambda|^{2}} L_{n}(2\eta(t)|\lambda|^{2})\exp[-\big(1-\eta(t)+2\nu(t)\big)|\lambda|^{2}].
\end{align}
To obtain the Wigner function, rather than directly Fourier-transforming this expression, it is easier to introduce the generating function
\begin{align}
G_{\chi}(\lambda,\lambda^{*},t)&=\sum_{n=0}^{\infty}s^{n}\chi_{s}(\lambda,\lambda^{*},t),
\end{align}
so that one can use the generating function for the Laguerre polynomials
\begin{align}
\sum_{n=0}^{\infty}s^{n}L_{n}(x)&=\frac{1}{1-s}\exp\left(-\frac{xs}{1-s}\right).
\end{align}
By direct computation, we find
\begin{align}
G_{\chi}(\lambda,\lambda^{*},t)&=\exp\Big[-\Big(1+2\nu(t)\Big)|\lambda|^{2}\Big]\frac{1}{1-s}\exp\left(-\frac{2\eta(t)|\lambda|^{2}s}{1-s}\right)\\
&=\frac{1}{1-s}\exp\left[-\Big(1+2\nu(t)+\frac{2s}{1-s}\eta(t)\Big)|\lambda|^{2}\right].
\end{align}
The Fourier transform of the generating function is given by 
\begin{align}
G_{W}(\alpha,\alpha^{*},t)&=\frac{2}{\pi}\int d^{2}\lambda\;e^{\lambda \alpha^{*}-\lambda^{*}\alpha}G_{\chi}(\lambda,\lambda^{*},t).
\end{align}
However, we have the complex integral
\begin{align}
\frac{2}{\pi}\int d^{2}\lambda e^{-B|\lambda|^{2}}e^{-\lambda^{*}\alpha+\lambda\alpha^{*}}=\frac{2}{B}\exp\left(-\frac{|\alpha|^{2}}{B}\right).
\end{align}
Hence, 
\begin{align}
G_{W}(\alpha,\alpha^{*},t)&=\frac{2}{D+(2\eta-D)s}\exp\left[-\frac{|\alpha|^{2}(1-s)}{D+(2\eta-D)s}\right],
\end{align}
where we introduced $D=1+2\nu(t)$. Let us now use the identity
\begin{align}
\frac{1-s}{1-qs}&=1-\frac{(1-q)s}{1-qs}.
\end{align}
Hence, we have
\begin{align}
G_{W}(\alpha,\alpha^{*},t)&=\frac{2}{D+(2\eta-D)s}\exp\left(-\frac{|\alpha|^{2}}{D}\right)\exp\left(\frac{|\alpha|^{2}}{D}\frac{1-q}{q}\frac{y}{1-y}\right),\\
q&=\frac{D-2\eta}{D},\\
y&=qs.
\end{align}
Furthermore, by using the generating function of the Laguerre polynomials
\begin{align}
G_{W}(\alpha,\alpha^{*})&=\frac{2}{D+(2\eta-D)s}\exp\left(-\frac{|\alpha|^{2}}{D}\right)(1-qs)\sum_{n=0}^{\infty}y^{n}L_{n}\left(-\frac{|\alpha|^{2}}{D}\frac{1-q}{q}\right)\\
&=\frac{2}{D}\exp\left(-\frac{|\alpha|^{2}}{D}\right)\sum_{n=0}^{\infty}(qs)^{n}L_{n}\left(-\frac{|\alpha|^{2}}{D}\frac{2\eta}{D-2\eta}\right)\\
&=\sum_{n=0}^{\infty}s^{n}\left(\frac{D-2\eta}{D}\right)^{n}\frac{2}{D}\exp\left(-\frac{|\alpha|^{2}}{D}\right)L_{n}\left(-\frac{|\alpha|^{2}}{D}\frac{2\eta}{D-2\eta}\right).
\end{align}
Hence, the Wigner function of the $n$-th Fock state
\begin{align}
W(\alpha,\alpha^{*},t)&=\left(\frac{D-2\eta}{D}\right)^{n}\frac{2}{D}\exp\left(-\frac{|\alpha|^{2}}{D}\right)L_{n}\left(-\frac{|\alpha|^{2}}{D}\frac{2\eta}{D-2\eta}\right).
\end{align}
We can verify that this Wigner function is identical to the one obtained previously with Eq.~\eqref{Wigner_Fock_SU11}, by making the change of variables defined by the equations Eq.~\eqref{g0_nu}. We have
\begin{align}
1+g_{+}&=\frac{D}{1+\nu},& 1-g_{+}&=\frac{1}{1+\nu}.
\end{align}
Hence,
\begin{align}
\frac{1-g_{+}}{1+g_{+}}&=\frac{1}{D}.
\end{align}
Moreover, 
\begin{align}
\Lambda&=1-e^{-g_{0}}g_{-}(1+g_{+})=-\frac{(1+\nu)(D-2\eta)}{\eta}.
\end{align}
Note also
\begin{align}
(-1)^{n}\frac{\Lambda^{n}e^{n g_{0}}}{(1+g_{+})^{n}}
&=\left(\frac{D-2\eta}{D}\right)^{n},& \frac{2R}{(1+g_{+})\Lambda}&=-\frac{2 \eta R}{D(D-2\eta)}.
\end{align}
And
\begin{align}
\mathcal{N}e^{g_{0}/2}\frac{1}{1+g_{+}}=\frac{1}{D}.
\end{align}
Finally, we have
\begin{align}
W_{|n\rangle}(x,p,t)&=\mathcal{N}(t)\frac{1}{\pi}\frac{e^{g_{0}/2}}{(1+g_{+})}\exp\left(-R\frac{1-g_{+}}{1+g_{+}}\right)\frac{(-1)^{n} e^{ng_{0}}}{(1+g_{+})^{n}}\Lambda^{n}L_{n}\left[\frac{2 R }{(1+g_{+})\Lambda}\right]\\
&=\frac{1}{\pi D}\exp\left(-\frac{R}{D}\right)\left(\frac{D-2\eta}{D}\right)^{n}L_{n}\left[-\frac{2\eta R}{D(D-2\eta)}\right],
\end{align}
with $R=|\alpha|^{2}=x^{2}+p^{2}$. Hence, the two Wigner functions are related through the normalization factor
\begin{align}
W(\alpha,\alpha^{*},t)&=2\pi W_{|n\rangle}(x,p,t).
\end{align}
This is due to the difference in normalization
\begin{align}
\int dx\;dp\;W(x,p,t)=\int d^{2}\alpha\;\frac{1}{2\pi}W(\alpha,\alpha^{*},t)&=1.
\end{align}
This verifies that the Wigner function agrees with the result obtained previously from the normal-ordered $su(1,1)$ evolution, after accounting for the different normalization conventions used in the two phase-space representations.
\section{Binomial code-space survival probability}
\label{app:binomial_correlations}
In this section, we illustrate the use of the exact propagator $V(t,0)$ by evaluating the survival probability of a finite-support binomial bosonic code. Consider the error set
\begin{align}
\overline{\varepsilon}&=\{\mathbbm{1},a,a^{2},\cdots,a^{L},a^{\dagger},\cdots,(a^{\dagger})^{G},n,\cdots, n^{D}\},
\end{align}
which includes up to $L$ photon losses, up to $G$ photon gain errors, and up to $D$ dephasing events. The simple class of codes which can correct such an error set introduced in \cite{Michael2016} is
\begin{align}
|W_{\uparrow/\downarrow}\rangle&=\frac{1}{\sqrt{2^{N}}}\sum_{p\quad\text{even/odd}}^{[0,N+1]}\sqrt{\binom{N+1}{p}}|p(S+1)\rangle,
\end{align}
where the spacing is $S=L+G$, maximum order $N={\rm max}\{L,G,2D\}$ and the range of the index $p$ is from $0$ to $N+1$. Hence, for the choice of parameters $L=G=1$ and $D=0$, one obtains the encoding
\begin{align}
|0_{L}\rangle&=\frac{|0\rangle+|6\rangle}{\sqrt{2}},& |1_{L}\rangle&=|3\rangle.
\end{align}
We are interested in characterizing the evolution of this code under gain--loss dynamics. The equation of evolution of the off-diagonal Fock matrix element is given by Eq.~\eqref{eq:transition_kernel_definition_main} in terms of the transition kernel Eq.~\eqref{eq:transition_kernel}. Equation~\eqref{eq:transition_kernel} implies the selection rule
\begin{align}
T_{rq|mn}(t,0)
&=
0
\quad\text{if}\quad
r-q\neq m-n.
\label{eq:coherence_order_selection}
\end{align}
The gain--loss channel therefore preserves the Fock-space coherences such that
$r-q=m-n$. The density matrix elements at time $t$ are given by
\begin{align}
\rho_{kn}(t)
&=
\sum_{a,b=0}^{\infty}
T_{kn|ab}(t,0)
\rho_{ab}(0).
\label{eq:rho_first_time}
\end{align}
This leads to the general expression for the survival probability in the main text  Eq.~\eqref{survival_probability}. To characterize the channel independently of a preferred logical
direction, we take the initial state to be maximally mixed within the code
space, Eq.~\eqref{eq:binomial_initial_state} in the main text. Let us now detail the expression for the different kernels in the expression of the code-space survival probability Eq.~\eqref{survival_probability}. The initial projector in the Fock basis is
\begin{align}
P_{C}&=\frac{1}{2}\big(|0\rangle\langle 0|+|0\rangle\langle 6|+|6\rangle\langle 0|+|6\rangle\langle 6|\big)+|3\rangle\langle 3|.
\end{align}
By the selection rule, we have (in making explicit only the terms that will contribute to the projection $P_{C}$)
\begin{align}
V(t,0)(|0\rangle\langle 0|)&=T_{66|00}|6\rangle\langle 6|+T_{00|00}|0\rangle\langle 0|+T_{33|00}|3\rangle\langle 3|+\cdots,\\
V(t,0)(|0\rangle\langle 6|)&=T_{06|06}|0\rangle\langle 6|+\cdots,\\
V(t,0)(|6\rangle\langle 0|)&=T_{60|60}|6\rangle\langle 0|+\cdots,\\
V(t,0)(|3\rangle\langle 3|)&=T_{33|33}|3\rangle\langle 3|+T_{00|33}|0\rangle\langle 0|+T_{66|33}|6\rangle\langle 6|+\cdots,\\
V(t,0)(|6\rangle\langle 6|)&=T_{00|66}|0\rangle\langle 0|+T_{66|66}|6\rangle\langle 6|+T_{33|66}|3\rangle\langle 3|+\cdots
\end{align}
One obtains
\begin{align}
V(t,0)[\rho(0)]&=\frac{1}{4}(T_{66|00}|6\rangle\langle 6|+T_{00|00}|0\rangle\langle 0|+T_{33|00}|3\rangle\langle 3|+T_{06|06}|0\rangle\langle 6|+T_{60|60}|6\rangle\langle 0|\nonumber\\
&+T_{00|66}|0\rangle \langle 0|+T_{66|66}|6\rangle\langle 6|+T_{33|66}|3\rangle\langle 3|)\nonumber\\
&+\frac{1}{2}(T_{33|33}|3\rangle\langle 3|+T_{00|33}|0\rangle\langle 0|+T_{66|33}|6\rangle\langle 6|)+\cdots,
\end{align}
Hence, we obtain
\begin{align}
{\rm tr}[P_{C}V(t,0)\rho(0)]&=\frac{1}{8}(T_{66|00}+T_{00|00}+2T_{33|00}+T_{06|06}+T_{60|60}+T_{00|66}+T_{66|66}+2T_{33|66})\nonumber\\
&+\frac{1}{4}(2T_{33|33}+T_{00|33}+T_{66|33}).
\end{align}
To make the expression more readable, we introduce the notations
\begin{align}
\zeta&=e^{g_{0}/2},& \mu&=g_{-}, & \vartheta&=g_{+}.
\end{align}
Hence, we obtain the following matrix elements
\begin{align}
T_{00|00}&=\mathcal{N}\zeta,\\
T_{33|00}&=\mathcal{N}\zeta \vartheta^{3},\\
T_{66|00}&=\mathcal{N}\zeta \vartheta^{6},\\
T_{00|33}&=\mathcal{N}\zeta  \mu^{3},\\
T_{33|33}&=\mathcal{N}(\zeta \mu^{3}\vartheta^{3}+9 \mu^{2}\vartheta^{2}\zeta^{3}+9 \mu \vartheta \zeta^{5}+\zeta^{7}),\\
T_{66|33}&=\mathcal{N}(\zeta \mu^{3}\vartheta^{6}+18 \mu^{2}\vartheta^{5}\zeta^{3}+45 \mu \vartheta^{4}\zeta^{5}+20 \vartheta^{3}\zeta^{7}),\\
T_{00|66}&=\mathcal{N}\mu^{6}\zeta,\\
T_{33|66}&=\mathcal{N}(\zeta \mu^{6}\vartheta^{3}+18 \mu^{5}\vartheta^{2}\zeta^{3}+45 \mu^{4}\vartheta \zeta^{5}+20 \mu^{3}\zeta^{7}),\\
T_{66|66}&=\mathcal{N}(\zeta \mu^{6}\vartheta^{6}+36 \mu^{5}\vartheta^{5}\zeta^{3}+225 \mu^{4}\vartheta^{4}\zeta^{5}+400 \mu^{3}\vartheta^{3}\zeta^{7}+225 \mu^{2}\vartheta^{2}\zeta^{9}+36 \mu \vartheta \zeta^{11}+\zeta^{13}),\\
T_{06|06}&=\mathcal{N}\zeta^{7},\\
T_{60|60}&=\mathcal{N}\zeta^{7}.
\end{align}
Collecting these terms, we obtain the expression for the survival probability in the main text Eq.~\eqref{analytical_pC}.
\end{document}